\documentclass[aps,prd,showpacs,eqsecnum,twocolumn,superscriptaddress,a4paper,nofootinbib]
{revtex4-1}
\usepackage{amsmath,amssymb,graphicx,color}
\usepackage{bm}
\usepackage{epsf}
\newcommand{\ms}[1]{\textcolor{black}{#1}}
\newcommand{\ez}[1]{\textcolor{black}{#1}}

\begin{document}

\title{Constraint on the maximum mass of neutron stars using GW170817 event
}

\author{Masaru Shibata} 
\affiliation{Max Planck Institute for
  Gravitational Physics (Albert Einstein Institute), Am Mühlenberg 1,
  Potsdam-Golm 14476, Germany}
\affiliation{Yukawa Institute for Theoretical
  Physics, Kyoto University, Kyoto, 606-8502, Japan~}
\author{Enping Zhou} 
\affiliation{Max Planck Institute for
  Gravitational Physics (Albert Einstein Institute), Am Mühlenberg 1,
  Potsdam-Golm 14476, Germany}

\author{Kenta Kiuchi}
\affiliation{Max Planck Institute for
  Gravitational Physics (Albert Einstein Institute), Am Mühlenberg 1,
  Potsdam-Golm 14476, Germany}
\affiliation{Yukawa Institute for Theoretical
  Physics, Kyoto University, Kyoto, 606-8502, Japan~}

\author{Sho Fujibayashi} 
\affiliation{Max Planck Institute for
  Gravitational Physics (Albert Einstein Institute), Am Mühlenberg 1,
  Potsdam-Golm 14476, Germany}


\date{\today}
\newcommand{\beq}{\begin{equation}}
\newcommand{\eeq}{\end{equation}}
\newcommand{\beqn}{\begin{eqnarray}}
\newcommand{\eeqn}{\end{eqnarray}}
\newcommand{\pa}{\partial}
\newcommand{\vp}{\varphi}
\newcommand{\varep}{\varepsilon}
\newcommand{\ep}{\epsilon}
\newcommand{\comp}{(M/R)_\infty}
\begin{abstract}

We revisit the constraint on the maximum mass of cold spherical
neutron stars coming from the observational results of GW170817.  We
develop a new framework for the analysis by employing both energy and
angular momentum conservation laws as well as solid results of latest
numerical-relativity simulations and of neutron stars in equilibrium.
The new analysis shows that the maximum mass of cold spherical neutron
stars can be only weakly constrained as $M_{\rm max} \alt 2.3M_\odot$.
\ms{Our present result illustrates that the merger remnant neutron
  star at the onset of collapse to a black hole is not necessarily
  rapidly rotating and shows that we have to take into account the
  angular momentum conservation law to impose the constraint on the
  maximum mass of neutron stars.}

\end{abstract}


\maketitle


\section{Introduction}

The first direct detection of gravitational waves from the coalescence
of binary neutron stars (GW170817)~\cite{GW170817} was accompanied
with a wide variety of the observations of electromagnetic
counterparts~\cite{GW170817a}. These observations give us new
constraints for the properties of neutron stars.  Gravitational-wave
observation for the late inspiral phase of the binary neutron stars
constrains the binary tidal deformability in the range of $100 \alt
\Lambda \alt 800$~\cite{GW170817b,De}. This suggests that the radius
of the $1.4M_\odot$ neutron star would be in the range between $\sim
10.5$\,km and $\sim 13.5$\,km.

Electromagnetic observations, in particular a possible observation of
a gamma-ray burst~\cite{GW170817c} and ultraviolet-optical-infrared
observation~\cite{GW170817a,EM2017}, are also used for constraining
the maximum mass of neutron stars. If we assume that the gamma-ray
burst was driven from a remnant black hole surrounded by a torus, the
black hole might have to be formed in a short timescale after the
merger. In this hypothesis, Ref.~\cite{Rezzolla18} suggests that the
maximum mass of neutron stars would be fairly small central value of
$\approx 2.17M_\odot$.  Reference~\cite{Ruiz18} suggests that the
maximum mass of neutron stars would be 2.16--$2.28M_\odot$, supposing
that the merger remnant neutron star has the same mass as that of the
inspiraling binary neutron star and is a rapidly rotating state (note
that the actual mass of the remnant neutron star should be smaller:
see Sec.~II).  The optical and infrared counterparts also suggest that
the remnant massive neutron star would survive at least for several
hundreds ms~\cite{MM2017,shibata17,Albino,Lippuner}, while the absence
of an extremely bright emission in the ultraviolet and optical bands
at $\alt 1$\,d suggests that the remnant formed after the merger would
collapse to a black hole within a timescale of $\sim 100$\,s after the
merger~\cite{MP14,MM2017,shibata17}.  This speculation could also give
a constraint on the maximum mass of neutron stars, and
Refs~\cite{MM2017,shibata17} suggest a fairly small value of the
maximum mass as $\alt 2.2M_\odot$.  However, these constraints are
imposed in the assumption that the merger remnant neutron star is
rapidly rotating at the onset of collapse; the constraint is imposed
without taking into account details of angular momentum dissipation
process in the post-merger stage self-consistently.

In this paper, we revisit the constraint on the maximum mass of cold
spherical neutron stars imposed by the observational results of
GW170817. The analysis is done taking into account the evolution
process in the post-merger phase of binary neutron stars, by carefully
analyzing both energy and angular momentum conservation laws and by
employing solid results of latest numerical-relativity simulations and
of rotating neutron stars in equilibrium. In particular, we show that
it is essential to take into account the angular momentum conservation
for this kind of analysis.  We then find that the maximum mass of cold
spherical neutron stars could be as large as $\sim 2.3M_\odot$: that
is, \ms{the upper bound may be by $\sim 0.1M_\odot$ larger than in our
  previous analysis~\cite{shibata17}}. We thus conclude that a
simplified analysis and an inappropriate assumption lead to an
inaccurate constraint on the maximum mass.

The paper is organized as follows. In Sec.~\ref{sec2}, we describe 
assumptions imposed in our present analysis and resulting basic
equations. In Sec.~\ref{sec3}, we derive the constraint on the
maximum mass of cold spherical neutron stars.  Section~\ref{sec4} is
devoted to a summary.  Throughout this paper, $G$ and $c$ denote the
gravitational constant and speed of light, respectively. 

\section{Assumptions and basic equations}\label{sec2}

\begin{figure}[t]
\epsfxsize=3.4in \leavevmode \epsffile{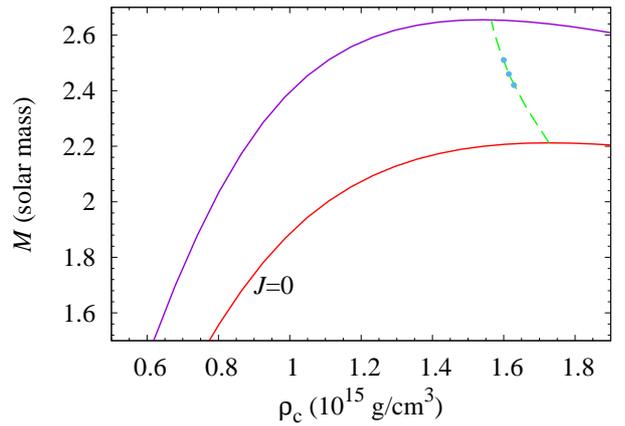}
\caption{Several important curves for rigidly rotating neutron stars
  in the plane of the gravitational mass ($M$) as a function of the
  central density ($\rho_{\rm c}$). In this example, we employ EOS-6
  (see Sec.~\ref{sec3}) as the neutron-star equation of state.  The
  lower and upper solid curves show the sequences of non-rotating
  neutron stars (labeled by $J=0$) and rigidly rotating neutron stars
  at mass shedding limits.  The dashed curve shows the sequence of
  neutron stars that are marginally stable to gravitational collapse
  (the sequence of the turning points); the neutron stars in the
  lower-density side of this dashed curve are stable, otherwise they
  are unstable.  The filled circles denote the predicted points at the
  onset of collapse for GW170817 in this model equation of state with
  $M_{\rm out}+M_{\rm eje}=0.048$, 0.096, and $0.150M_\odot$ (from
  higher to lower mass: see Sec.~\ref{sec3}).
\label{fig0}}
\end{figure}

In this paper, we postulate or assume the following interpretation for
the observational results of neutron-star merger event GW170817.
First, we postulate that a remnant neutron star was formed after the
merger and \ms{survived temporarily as a quasi-steady strong neutrino
  emitter}. This is supported by the observations of electromagnetic
(ultraviolet-optical-infrared) counterparts for the merger
event~\cite{EM2017}, because the neutrino irradiation from the remnant
neutron stars would play an important role for reducing the neutron
richness and lanthanide fraction of the ejecta
(e.g.,~Refs.~\cite{EM2017,shibata17,Albino,Lippuner}).  Second, we
postulate that the remnant neutron star collapsed to a black hole
within the dissipation timescale of its kinetic energy via
electromagnetic radiation like the magnetic dipole radiation (i.e.,
within $\sim 100$\,sec). This is because after the merger of binary
neutron stars, magnetic fields are likely to be significantly
amplified in the remnant neutron star~\cite{PR} and in the presence of
a strong energy injection comparable to the rest-mass energy of the
ejecta by the electromagnetic radiation associated with large kinetic
energy of the remnant, the ultra-violate-optical-infrared counterparts
for GW170817 would be much brighter than the observational
results~\cite{MM2017,shibata17}.  However, it should be noted that if
the rotational kinetic energy of the remnant neutron star is
dissipated by gravitational radiation and/or carried away by mass
ejection in a short timescale (e.g., $\sim 10$\,s), we may accept the
formation of a stable neutron star, although this possibility is not
very likely (see Sec.~\ref{sec3f} for a discussion).  Third, we {\em
  assume} that the remnant neutron star at the onset of collapse was
rigidly rotating. This could be a reasonable assumption for the case
that the remnant neutron star was long-lived, because the degree of
the differential rotation is likely to be reduced sufficiently via
long-term angular momentum transport process (e.g.,
Ref.~\cite{Fujiba2018}).  Thus, we suppose that the collapse of the
remnant neutron star should occur at a turning point (a marginally
stable state) along an equilibrium sequence of rigidly rotating
supramassive neutron stars~\cite{FIS,CST94} (see Fig.~\ref{fig0} for
turning points).  \ms{Fourth}, we postulate that an appreciable
fraction of the baryon of mass $\agt 0.05M_\odot$ was located outside
the black hole at its formation because the
ultraviolet-optical-infrared observations for the GW170817 event
indicate that the ejecta mass was likely to be $\agt
0.03M_\odot$~\cite{EM2017}.  \ms{Finally, in this paper, we assume
  that neutron stars are described by simple nuclear matter equations
  of state~(see Appendix A) and do not suppose other exotic
  possibilities like quark stars, twin stars, and hybrid stars~(e.g.,
  Ref.~\cite{Exotic}). We also postulate that general relativity is a
  high-precision theory for neutron stars and binary neutron star
  mergers.}

The quantities which are referred to in this paper are as
follows; the baryon rest mass and gravitational mass at the formation
of the binary neutron stars, $M_*$ and $M$, respectively; the torus
mass around the remnant neutron star and ejecta mass at the formation of a 
black hole, $M_{\rm out}$ and $M_{\rm eje}$, respectively; the total energy
radiated by gravitational waves throughout the inspiral to 
post-merger phases, $E_{\rm GW}$; total energy radiated by neutrinos
throughout the merger to post-merger phases, $E_\nu$; and the maximum
mass for cold spherical neutron stars, $M_{\rm max}$.
Here, we know that $M=2.74^{+0.04}_{-0.01}M_\odot$ with the 90\%
credible level~\cite{GW170817,GW170817b}.  In $M_{\rm out}$, we
include mass of an atmosphere surrounding the central object.
$E_{\rm GW}$ can be divided into two parts: One is that emitted in the
inspiral phase, $E_{\rm GW,i}$, and the other is that in the merger
and post-merger phases, $E_{\rm GW,p}$.  The subject of this paper is
to constrain the value of $M_{\rm max}$ by using the relations
satisfied among these quantities.

The first relation employed is the energy conservation law together
with the rest-mass conservation law, which gives
\beqn
M_{*f}:=M_*-M_{\rm out}-M_{\rm eje}=f_{\rm MS}M_f, \label{eq1.10}
\eeqn
where
\beqn
M_f:=M-E_{\rm GW}c^{-2}-E_\nu c^{-2}-M_{\rm out}-M_{\rm eje},~~
\label{eq1.1}
\eeqn
and $M_{*f}$ and $M_f$ denote the rest mass and gravitational mass at
the onset of collapse of the remnant neutron star, respectively.  In
Eq.~(\ref{eq1.1}), we assumed that the thermal, kinetic, and
gravitational binding energy of the matter outside the remnant neutron
star is much smaller than (at most $\sim 10$\% of) its rest-mass
energy. This could affect the values of $M_f$, $M_{\rm out}$, and
$M_{\rm eje}$ only by $\sim 0.01M_\odot$.  We ignore the dissipation
by electromagnetic radiation like the magnetic dipole radiation
because in a short timescale of $\alt 10$\,s which we consider here,
the effect is likely to be negligible.

In Eq.~(\ref{eq1.10}), we denoted the ratio of the baryon rest mass to
the gravitational mass for the remnant neutron star at the onset of
collapse by $f_{\rm MS}\,(:=M_{*f}/M_f)$. For cold spherical neutron
stars at a marginally stable state to collapse, the ratio of the
baryon rest mass to the gravitational mass is $\approx 1.19 \pm 0.05$
(see Appendix A).  For rigidly rotating neutron stars near the
marginally stable state, this value is smaller by a factor of up to 
$\sim 0.02$, and thus, it depends only weakly on angular momentum
(cf.~Fig.~\ref{fig1} of Sec.~\ref{sec3}).

We also denote the ratio of $M_*$ to $M$ as $f_0:=M_*/M$.  Since the
ratio of the baryon rest mass to the gravitational mass is $\sim
1.08$--1.14 for realistic neutron stars with mass $1.2$--$1.6M_\odot$,
$f_0$ should be also $\approx 1.11 \pm 0.03$ (see Appendix A).

For larger radii of neutron stars, $f_0$ and $f_{\rm MS}$ are in
general smaller. It should be cautioned that the values of $f_0$ and
$f_{\rm MS}$ depend on the equation of state and their deviation is
not negligible for imposing the constraint to $M_{\rm max}$ \ms{within
  an error of $0.1M_\odot$}. Thus, in the analysis for constraining
$M_{\rm max}$, we must not use particular values for them.

By eliminating $M_*$ from Eqs.~(\ref{eq1.10}) and (\ref{eq1.1}), we
finally obtain
\beqn
M\left(1-{f_0 \over f_{\rm MS}}\right)
&=&E_{\rm GW}c^{-2}+E_\nu c^{-2} \nonumber \\
&&+(M_{\rm out}+M_{\rm eje})\left(1-{1 \over f_{\rm MS}}\right).~~
\label{eq1.2}
\eeqn
From Eqs.~(\ref{eq1.1}) and (\ref{eq1.2}), we also obtain the gravitational 
mass at the onset of collapse as
\beqn 
M_f ={f_0 \over f_{\rm MS}}M - {M_{\rm out}+M_{\rm eje} \over
  f_{\rm MS}}.
\label{eq1.3}
\eeqn
As we show in Appendix A, for cold spherical neutron stars of a
variety of equations of state with the constraint that the tidal
deformability of $1.35M_\odot$ neutron stars, $\Lambda_{1.35}$, is
smaller than $1000$, the value of $f_0/f_{\rm MS}$ is approximately
$0.920 \pm 0.025$. For this range, the left-hand side of
Eq.~(\ref{eq1.2}) is found to be in a wide range as
\beq
M\left(1-{f_0 \over f_{\rm MS}}\right)=0.219^{+0.073}_{-0.069}M_\odot, 
\label{eq1.30}
\eeq
and also, 
\beq
M_f=2.521^{+0.106}_{-0.077}M_\odot-{M_{\rm out}+M_{\rm eje} \over f_{\rm MS}}.
\label{eq1.31}
\eeq
We note that $f_{\rm MS}$ depends only weakly on the angular momentum,
as we find in Sec.~\ref{sec3}.  Equations~(\ref{eq1.30}) and
(\ref{eq1.31}) clearly illustrate that the dissipated energy and the
gravitational mass at the onset of collapse have uncertain of
approximately $\pm 0.1M_\odot$ because the equation of state is not
well constrained.  As we show in Appendix A, the values of $f_0/f_{\rm
  MS}$ are correlated with $M_{\rm max}^{-1}$.  Thus, smaller values
of $M_{\rm max}$ lead to larger values of $M_f$ and to smaller energy
dissipated from the system (see Sec.~\ref{sec3}).

The electromagnetic (ultraviolet-optical-infrared) counterpart
observations for GW170817 show that the ejecta mass would be
approximately $M_{\rm eje}\approx 0.03$--$0.05M_{\odot}$~(e.g.,
Ref.~\cite{EM2017}).  This value is also supported by numerical
simulations for a neutron star surrounded by a
torus~\cite{MF2014,Perego14,Fujiba2018}. Numerical simulations (e.g.,
Refs.~\cite{shibata17,Fujiba2018,Fujiba2019}) also indicate that the
mass of the torus around the remnant neutron star would be $\sim
0.1$--$0.2M_\odot$ for its early stage, and reduce to $\sim
0.02$--$0.05M_\odot$ for its late stage $\tau \agt 1$\,s \ms{due to
  the mass ejection and mass accretion onto the remnant neutron star}.
Thus, $M_{\rm out}$ depends on the lifetime, $\tau$, of the remnant
neutron star.  Also, at the formation of a black hole, it is decreased
by $\agt 50\%$ because a substantial fraction of the torus matter in
the vicinity of the central object is swallowed by the black hole at
its formation. Thus, we suppose $0.02M_\odot \leq M_{\rm out} \leq
0.10M_\odot$ at the collapse of the remnant neutron star in the
following.  For $M_{\rm out}=0.06 \pm 0.04M_\odot$, $(M_{\rm
  out}+M_{\rm eje})/f_{\rm MS}$ is approximately $(0.083 \pm
0.042)M_\odot$ and $(M_{\rm out}+M_{\rm eje})(1-1/f_{\rm MS})$ is
$(0.016 \pm 0.008)M_{\odot}$, which is much smaller than the value of
Eq.~(\ref{eq1.30}). Using this result, we approximately obtain
\beqn
M_f=2.44^{+0.15}_{-0.12}M_\odot, 
\eeqn
and
\beqn
&&(E_{\rm GW}+E_\nu)c^{-2} \nonumber \\
&=&M\left(1-{f_0 \over f_{\rm MS}}\right)
- (M_{\rm out}+M_{\rm eje})\left(1-{1 \over f_{\rm MS}}\right) \nonumber \\
&=& (0.20 \pm 0.08)M_\odot. \label{egwenu}
\eeqn
Here the large (small) side of the uncertainty in $M_f$ comes basically from
the large (small) values of $f_0/f_{\rm MS}$, i.e., for small (large)
values of $E_{\rm GW}+ E_\nu$.  Taking into account the uncertainty in
these unknown values enlarges the uncertainty in the estimate of $M_f$
compared with our previous estimation~\cite{shibata17}.  It should be
also mentioned that the central value of $M_f$ is by $\sim
0.15M_\odot$ smaller than the value estimated in our previous
paper~\cite{shibata17}.  The reason for this is that we underestimated
the values of $E_{\rm GW}+E_\nu$ in the previous paper. 

$(E_{\rm GW}+E_\nu)c^{-2}$ is found to be typically $\sim 0.2M_\odot$
and at least larger than $\approx 0.12M_\odot$.  Numerical relativity
simulations~(e.g., Ref.~\cite{Zappa}) have shown that $E_{\rm GW,i}$
is well constrained to be $0.035$--$0.045M_\odot c^2$ for
$\Lambda_{1.35} \leq 1000$.  On the other hand, they show that $E_{\rm
  GW,p}$ depends strongly on the equation of state. It could be $\sim
0.125M_\odot c^2 \approx 2.5 \times 10^{53}$\,erg for the maximum
case~\cite{Zappa}. Note that $E_{\rm GW,p}$ should be smaller than the
rotational kinetic energy of the remnant neutron star (see also
Table~\ref{table:EOS2} of Sec.~\ref{sec3}),
\beqn
T={1 \over 2} I \Omega^2 &\approx& 2.3 \times 10^{53}\,{\rm erg}
\left({M_{\rm MNS} \over 2.6M_\odot}\right)
\left({R_{\rm MNS} \over 15\,{\rm km}}\right)^2
\nonumber \\
&& ~~~~~~~~~~~\times
\left({\Omega \over 10^4\,{\rm rad/s}}\right)^2, \label{eq1.32}
\eeqn
where $M_{\rm MNS}$, $R_{\rm MNS}$, and $\Omega$ are the gravitational
mass, equatorial circumferential radius, and angular velocity of the
remnant neutron star, and we assumed that it is rigidly rotating (and
hence its angular momentum is written as $2T/\Omega$). We note that
the maximum angular velocity is approximately written as (see also
Table~\ref{table:EOS2})
\beqn
\Omega_K &\approx &\sqrt{GM_{\rm MNS}/R_{\rm MNS}^3} \nonumber \\
&\approx& 1.01 \times 10^4\,{\rm rad/s} 
\left({M_{\rm MNS} \over 2.6M_\odot}\right)^{1/2}
\left({R_{\rm MNS} \over 15\,{\rm km}}\right)^{-3/2}. \nonumber \\
\eeqn

As we find in the following, one of the key parameters for determining
the angular momentum at the onset of collapse of the remnant neutron
star is $E_{\rm GW,p}$.  While $E_{\rm GW,p}$ could be $\sim
0.125M_\odot c^2$ if most of the rotational kinetic energy is
dissipated by gravitational radiation, the post-merger evolution
process is highly uncertain.  If an efficient angular momentum
transport works in the remnant neutron star and the degree of its
non-axisymmetric deformation is reduced quickly, $E_{\rm GW,p}$ would
be of order $0.01M_\odot c^2$~\cite{SK17}. These facts together with
Eq.~(\ref{eq1.30}) suggest that an appreciable amount of energy of
$\sim 0.1M_\odot c^2 \approx 2 \times 10^{53}$\,erg would be
dissipated by the emission of neutrinos until the onset of collapse to
a black hole, unless the remnant neutron star radiates gravitational
waves of energy comparable to $T$. Note that the value of $E_\nu \sim
10^{53}$\,erg is quite natural if the remnant neutron star is
long-lived with its lifetime $\tau \sim 1$\,s because
numerical-relativity simulations have shown that the neutrino
luminosity from the remnant neutron star, $L_\nu$, is of order
$10^{53}$\,erg/s~(e.g., Refs.~\cite{Sekig11,Foucart16,Radice16}). By
contrast, if the remnant neutron star is relatively short-lived with
$\tau \sim 100$\,ms, $E_\nu$ would be of $O(10^{52}\,{\rm erg})$
(i.e., $\alt 0.01M_\odot c^2$) and for this case, we have to employ a
large value of $E_{\rm GW,p}$ to satisfy Eq.~(\ref{egwenu}) (see
Sec.~\ref{sec3} for more specific examples).


Since the remnant neutron star is rotating, the gravitational mass at
the onset of collapse has to be larger than the maximum mass, $M_{\rm
  max}$, for cold spherical neutron stars by a factor of
$f_r=M_f/M_{\rm max} > 1$. Here, the value of $f_r$ is determined at a
turning point along an equilibrium sequence of rigidly rotating
neutron stars.  If we could obtain the value of $f_r$ together with
$M_f$, we can determine the value of $M_{\rm max}$.  \ms{For the
  change of $f_r$ by $0.1$, $M_{\rm max}$ could be changed by $\approx
  0.2M_\odot$. Thus, for constraining $M_{\rm max}$, e.g., within the
  $0.1M_\odot$ error, we have to determine the value of $f_r$ within
  an error of $\sim 0.05$.}  For the maximally rotating neutron star
along the turning point sequence, the value of $f_r$ is known to be
$\sim 1.2$~\cite{CST94,FF} (see also Fig.~\ref{fig1} of
Sec.~\ref{sec3}) and this value has been often used for guessing the
maximum mass of cold spherical neutron
stars~\cite{MM2017,shibata17,Rezzolla18,Ruiz18}. However, the remnant
neutron star is not always rotating in such a high rotation speed as
shown in Sec.~\ref{sec3}: In setting $f_r \approx 1.2$, one would
assume a particular angular momentum of the remnant neutron star
neglecting the angular momentum conservation.  In addition, the
maximum value of $f_r$ depends on the equation of state (see
Fig.~\ref{fig1} of Sec.~\ref{sec3}). It should be cautioned again that
in the analysis for constraining the value of $M_{\rm max}$, we must
not a priori employ a particular value for $f_r$.

For inferring the angular momentum of the remnant neutron star at the
onset of collapse, we have to seriously analyze the dissipation of
angular momentum in the post-merger phase.  Let $J_0$ and $J_f$ be the
angular momentum at the onset of merger and at the onset of collapse
of the remnant neutron star to a black hole, respectively.  Then, we
obtain
\beq
J_f=J_0 - J_{\rm GW,p} - J_\nu - J_{\rm out}-J_{\rm eje}, \label{eq1.4}
\eeq
where $J_{\rm GW,p}$, $J_\nu$, $J_{\rm out}$, and $J_{\rm eje}$ are
the angular momentum carried away (after the merger) by gravitational
radiation, by neutrinos, angular momentum of the torus surrounding the
remnant black hole at its formation, and angular momentum of ejecta
(at the black-hole formation). In the following, we give or determine
these quantities based on the results of numerical-relativity
simulations.  We also note that by angular momentum transport
processes from the remnant neutron star to the surrounding matter, the
angular momentum of torus and ejecta in general increases with time 
in the post-merger phase. 

Since gravitational waves emitted in the post-merger phase are
dominated by a fundamental mode of its frequency
$f=2$--4\,kHz~\cite{Bauswein,Hotoke13}, $J_{\rm GW}$ is approximately
written as
\beqn
&& J_{\rm GW,p} \approx {E_{\rm GW,p} \over \pi f} \nonumber \\
&\approx& 9.5 \times 10^{48}\, {\rm erg\,s}
\left({E_{\rm GW,p} \over 0.05M_{\odot}c^2}\right)
\left({f \over 3.0\,{\rm kHz}}\right)^{-1}.~~~~\label{Jgwp}
\eeqn
\ms{Our latest numerical-relativity simulation confirms that this
  relation is satisfied within 1\% accuracy~\cite{Kiuchi19}.}  Here,
for the binaries of total mass $\approx 2.7M_\odot$, $f \approx 3.6$,
3.1, and 2.5\,kHz for $R_{1.60} \approx 11$, 12, and
13.5\,km~\cite{Bauswein,Hotoke13} with $R_{M}$ the radius of a
spherical neutron star of its gravitational mass $M$ (in units of
$M_\odot$).  Thus, for $R_{1.35} \approx R_{1.60} \alt 13.5$\,km (this
constraint was given from the observational result of the tidal
deformability of GW170817~\cite{GW170817,GW170817b}), $f \agt
2.5$\,kHz.  In this paper, we infer the value of $f$ by using the
relation of Eq.~(3) of Ref.~\cite{Bauswein}.

Since the angular momentum of neutrinos are dissipated due to the fact
that the emitter (remnant neutron star) is rotating, $J_\nu$ is
written approximately by $J_\nu \approx (2/3)c^{-2} R_{\rm MNS}^2
\Omega E_\nu$~\cite{BS1998}, and thus,
\beqn
J_\nu &\approx& 3.0 \times 10^{48}\, {\rm erg\,s}
\left({E_\nu \over 0.1M_{\odot}c^2}\right)
\left({R_{\rm MNS} \over 15\,{\rm km}}\right)^{2} \nonumber \\
&& ~~~~~~~~~~~~~~~~\times\left({\Omega \over 10^4\,{\rm rad/s}} \right). 
\label{Jnu}
\eeqn
We note that the value of $J_\nu$ described here agrees with the
results of a numerical-relativity simulation within a factor of
2~\cite{Fujiba2019}. Equation~(\ref{Jnu}), in comparison with
Eq.~(\ref{Jgwp}), illustrates that this is a non-negligible but minor
effect for dissipating the angular momentum \ms{(thus, 
the error would be also a minor effect)}. 

$J_{\rm out}$ is associated with the typical radius of the torus
surrounding the remnant neutron star (at the onset of
collapse). Denoting it by $R_{\rm out}$, it is approximated by $J_{\rm
  out} \approx M_{\rm out} \sqrt{GM_{\rm MNS}R_{\rm out}}$, and thus,
\beqn
J_{\rm out} &\approx& 5.8 \times 10^{48}\, {\rm erg\,s}
\left({M_{\rm out} \over 0.05M_{\odot}}\right)
\left({R_{\rm out} \over 100\,{\rm km}}\right)^{1/2} \nonumber \\
&& ~~~~~~~~~~~~~~~~\times \left({M_{\rm MNS} \over 2.6M_\odot}\right)^{1/2}. 
\label{Jout}
\eeqn
Here $R_{\rm out}$ would be fairly small $\sim 50$\,km in the early
evolution stage of the accretion torus. However, during its long-term
viscous evolution as well as angular momentum transport from the
remnant neutron star, the typical radius increases to be $\agt
100$\,km for $\tau \agt 300$\,ms~\cite{Fujiba2018} which shows $R_{\rm out}
\approx 40+100 (\tau/1\,{\rm s})^{1/2}$\,km for $\tau \alt 1.5$\,s and
for a longer term, $R_{\rm out}$ approaches $\sim 200$\,km. 

$J_{\rm eje}$ is associated with the location at which the mass
ejection occurs. Denoting the typical location by $R_{\rm eje}$, it is
approximated by $J_{\rm eje} \approx M_{\rm eje} \sqrt{GM_{\rm MNS}R_{\rm
    eje}}$, and thus, 
\beqn
J_{\rm eje} &\approx& 6.9 \times 10^{48}\, {\rm erg\,s}
\left({M_{\rm eje} \over 0.05M_{\odot}}\right)
\left({R_{\rm eje} \over 140\,{\rm km}}\right)^{1/2} \nonumber \\
&& ~~~~~~~~~~~~~~~~\times \left({M_{\rm MNS} \over 2.6M_\odot}\right)^{1/2}. 
\label{Jeje}
\eeqn
For long-lived remnant neutron stars which we consider in this paper,
mass ejection mainly occurs through the long-term viscous process in
the post-merger stage from an accretion
torus~\cite{MF2014,Perego14,Fujiba2018} with mass $\sim 0.05M_\odot$,
while for dynamical mass ejection that occurs at merger, the matter
would be ejected at $R_{\rm eje} \sim 30$\,km with mass $\alt
0.01M_\odot$.  (Note that in $M_{\rm eje}$ both contributions are
included.)  Thus, for $J_{\rm eje}$, only the post-merger mass
ejection could contribute dominantly to the angular momentum
loss. Since this mass ejection is driven from the torus, we simply set
$R_{\rm out}=R_{\rm eje}$ in this paper; that is, we employ a relation
as $J_{\rm out}+J_{\rm eje}=(M_{\rm out}+M_{\rm eje})\sqrt{GM_{\rm
    MNS}R_{\rm out}}$. Here, we may overestimate $J_{\rm eje}$ because
the dynamical ejecta would have smaller angular momentum than the
post-merger one. However, since the mass of the dynamical ejecta would
be much smaller than the post-merger ejecta, the degree of the
overestimation would be minor. 

\ms{We note that the error for the estimate in $J_{\rm out}+J_{\rm eje}$ 
associated with the uncertainty in $R_{\rm out}$ would be of order 
10\%. This error will be reflected in $M_{\rm out}+M_{\rm eje}$ 
for determining it (see Sec.~\ref{sec3}). However, its error 
size does not affect our final conclusion.}

Thus, besides $J_0$, the quantities in the right-hand side of
Eq.~(\ref{eq1.4}) is related to $E_{\rm GW}$, $E_\nu$, $M_{\rm out}$,
and $M_{\rm eje}$ with $f$, $\Omega$, $M_{\rm MNS}$, and $R_{\rm out}$ 
as given parameters. 

In addition, we have an important relation. From the definitions 
already shown, we obtain
\beqn
f_r M_{\rm max}=M_f
={f_0 \over f_{\rm MS}}M-{M_{\rm out}+M_{\rm eje} \over f_{\rm MS}}. 
\eeqn
Thus, for a given value of $f_r$ and $(M_{\rm out}+M_{\rm eje})/f_{\rm MS}$,
$f_0/f_{\rm MS}$ becomes a linear function of $M_{\rm max}$ as
\beqn
{f_0 \over f_{\rm MS}}= {M_{\rm out}+M_{\rm eje} \over f_{\rm MS}M}
+{f_r M_{\rm max} \over M}. \label{frrel}
\eeqn
This becomes a condition that determines a particular state of the
remnant neutron star at the onset of collapse for a given equation of
state. This equation plays an important role in Sec.~\ref{sec3}.

Note that for larger values of $M_f$, $E_{\rm GW,p}+E_\nu$ should be
smaller, and thus, the value of $f_r$ should be larger because the
collapse to a black hole should occur before a substantial fraction of
angular momentum (and energy) is dissipated.  Remembering the fact
that $M_f$ is correlated with $M_{\rm max}^{-1}$ for plausible
equations of state (see Appendix A), we then find that for the larger
value of $M_{\rm max}$, the required value of $f_r$ is smaller.  This
fact clarifies that we cannot a priori give the value of $f_r$ for
constraining $M_{\rm max}$.

\begin{table*}[t]
\caption{Selected piecewise polytropic equations of state and
  important quantities for spherical neutron stars.  The units of the
  mass and radius are $M_\odot$ and kilometer, and that of $p$ is
  ${\rm dyn/cm^2}$. $f_{\rm MS}$ is the ratio of the baryon rest mass
  to $M_{\rm max}$ for the maximum mass neutron star.  $f_0$ shown
  here is $M_*/M$ for binaries of mass $1.35M_\odot$ and
  $1.40M_\odot$. $f$ denotes the dominant frequency of post-merger
  gravitational waves predicted approximately by the formula in
  Ref.~\cite{Bauswein}.} 
 \begin{tabular}{ccccccccccc} \hline
~~~Model~~~ & ~~$\Gamma_2$~~ & ~~$\Gamma_3$~~ & ~$\log_{10} p$~ 
& ~~$M_{\rm max}$~~ & ~~$f_{\rm MS}$~~ & ~~$R_{1.60}$~~ & ~~$R_{1.35}$~~ 
& ~$\Lambda_{1.35}$~ & ~~~~$f_0$~~~~ & ~~$f$\,(kHz)~~
\\
 \hline \hline
EOS-1 &  3.15 & 2.81 & 34.350 & 2.075& 1.200 & 11.27& 11.30 & 366.6 & 1.113 & 3.45\\
EOS-2 &  2.60 & 2.84 & 34.550 & 2.106& 1.172 & 12.67& 12.94 & 746.0 & 1.092 & 2.71\\
EOS-3 &  3.45 & 2.70 & 34.300 & 2.113& 1.208 & 11.17& 11.12 & 348.2 & 1.117 & 3.50\\ 
EOS-4 &  3.80 & 2.80 & 34.200 & 2.147& 1.221 & 10.91& 10.80 & 302.8 & 1.122 & 3.62\\ 
EOS-5 &  2.70 & 2.78 & 34.575 & 2.176& 1.177 & 12.88& 13.06 & 821.9 & 1.092 & 2.65\\
EOS-6 &  3.00 & 2.80 & 34.500 & 2.212& 1.196 & 12.21& 12.25 & 599.0 & 1.102 & 3.01\\
EOS-7 &  3.15 & 2.81 & 34.475 & 2.246& 1.204 & 12.06& 12.04 & 555.7 & 1.105 & 3.08\\
EOS-8 &  3.65 & 2.78 & 34.325 & 2.252& 1.222 & 11.39& 11.27 & 395.2 & 1.116 & 3.40\\
EOS-9 &  3.05 & 2.80 & 34.550 & 2.306& 1.200 &12.57& 12.56 & 720.7 & 1.099 & 2.74\\
EOS-10 &  2.85 & 2.85 & 34.625 & 2.328& 1.189 & 13.24& 13.29 & 967.6 & 1.092 & 2.55\\
EOS-11&  3.80 & 2.50 & 34.375 & 2.353& 1.229 & 11.66& 11.50 & 459.7 & 1.113 & 3.27\\ 
EOS-12&  3.25 & 2.78 & 34.575 & 2.433& 1.212 & 12.68& 12.60 & 757.8 & 1.100 & 2.70\\
 \hline
 \end{tabular}
 \label{table:EOS}
\end{table*}

\section{Analysis based on numerical modeling of neutron stars and binary neutron star 
mergers}\label{sec3}

\subsection{Preparation}

\begin{figure*}[t]
\epsfxsize=3.4in \leavevmode \epsffile{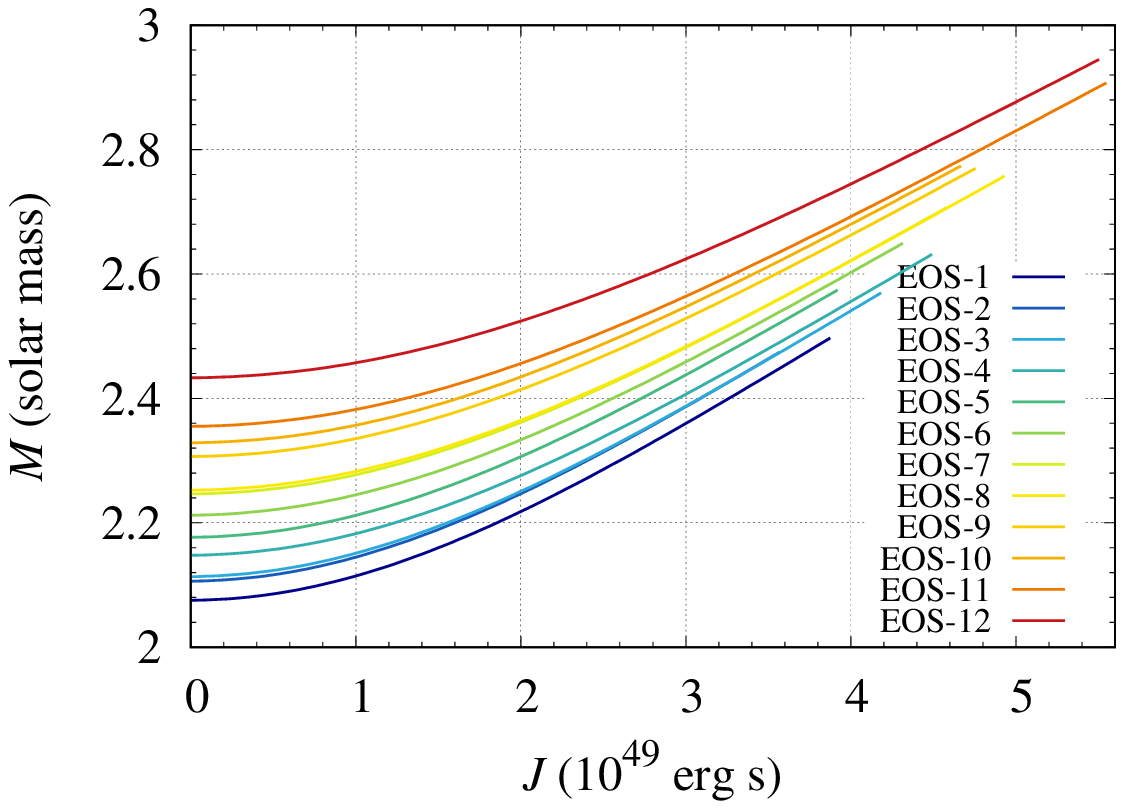}~~
\epsfxsize=3.4in \leavevmode \epsffile{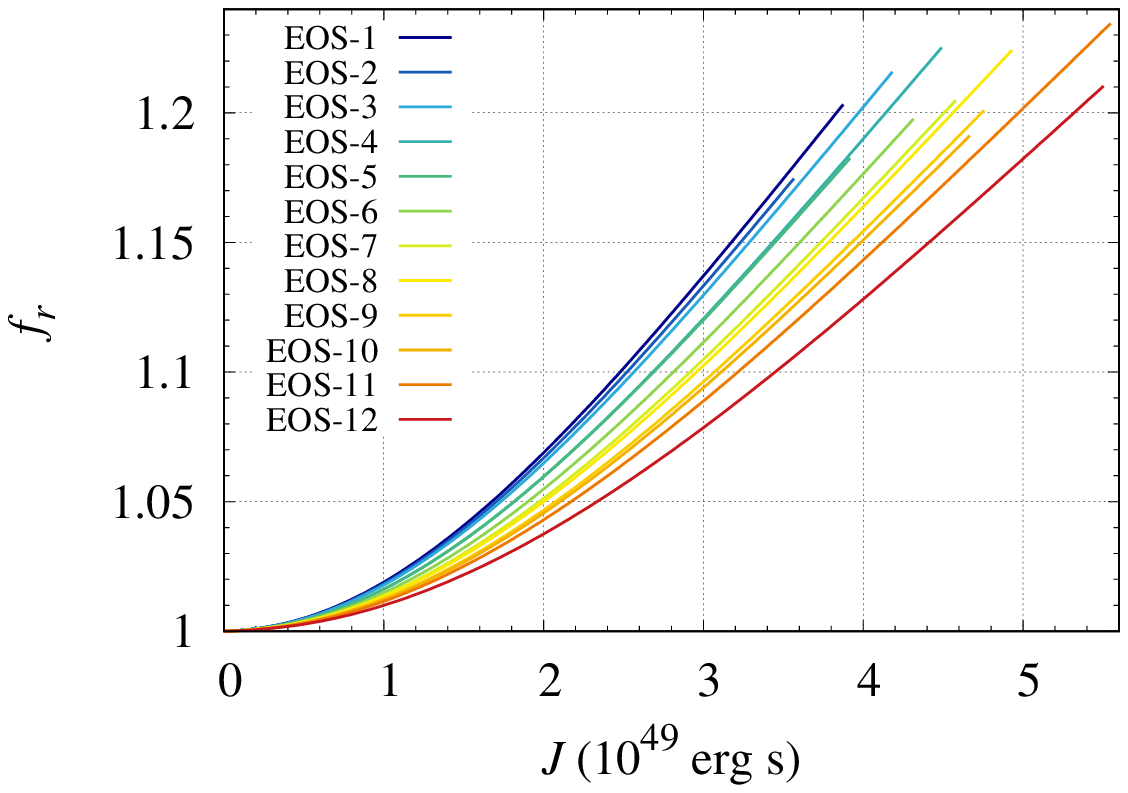}\\
\epsfxsize=3.4in \leavevmode \epsffile{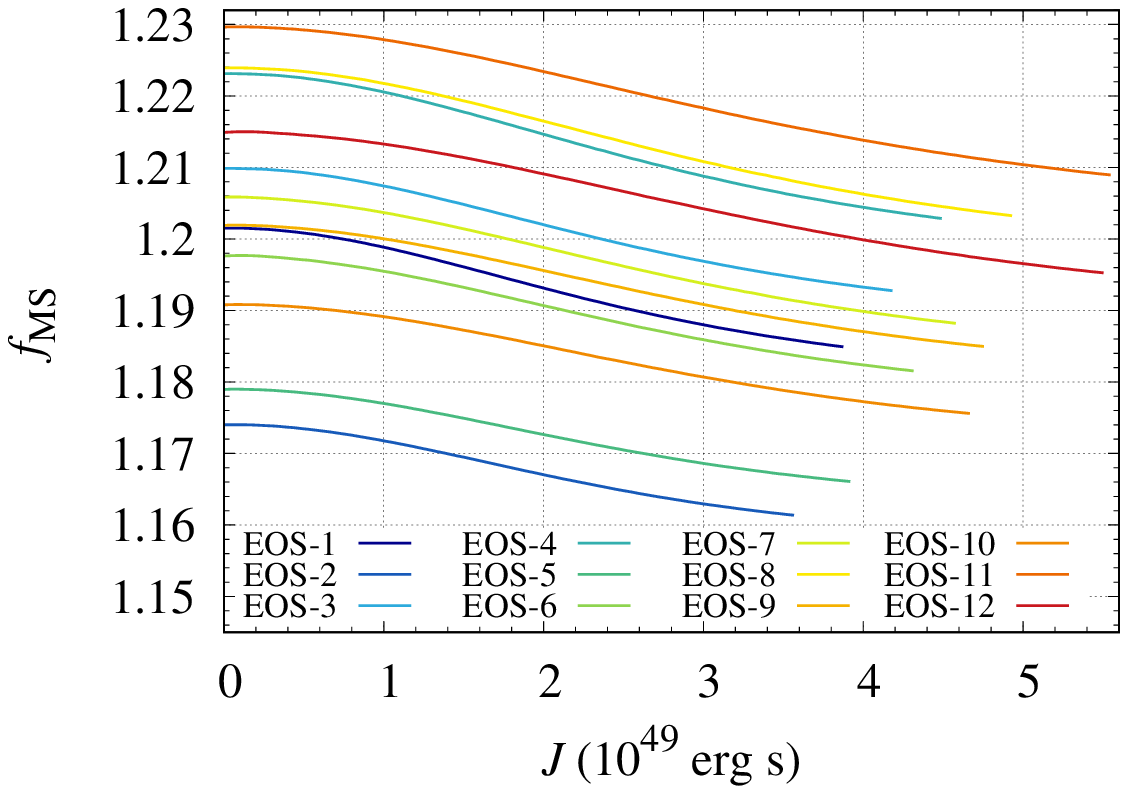}~~
\epsfxsize=3.4in \leavevmode \epsffile{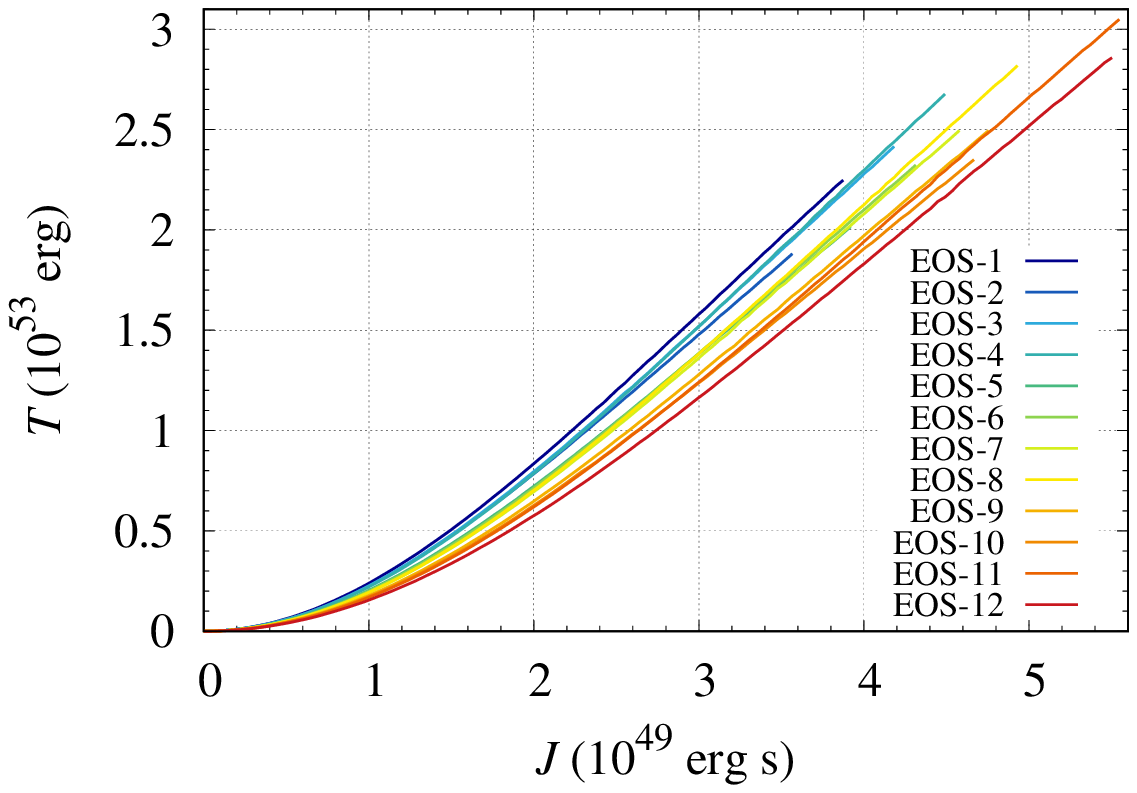}
\caption{Gravitational mass, $M$, ratio of $M$ to $M_{\rm max}$
  ($f_r$), ratio of baryon rest mass to the gravitational mass,
  $f_{\rm MS}$,
and rotational kinetic energy $T=J\Omega/2$ as functions of $J$ for
rigidly rotating neutron stars along the marginally stable sequence
(i.e., along the turning point sequence) with selected equations of
state (see Table~\ref{table:EOS}).  Note that $M$ at $J=0$ is $M_{\rm
  max}$.
\label{fig1}}
\end{figure*}

\begin{figure*}[t]
\epsfxsize=3.4in \leavevmode \epsffile{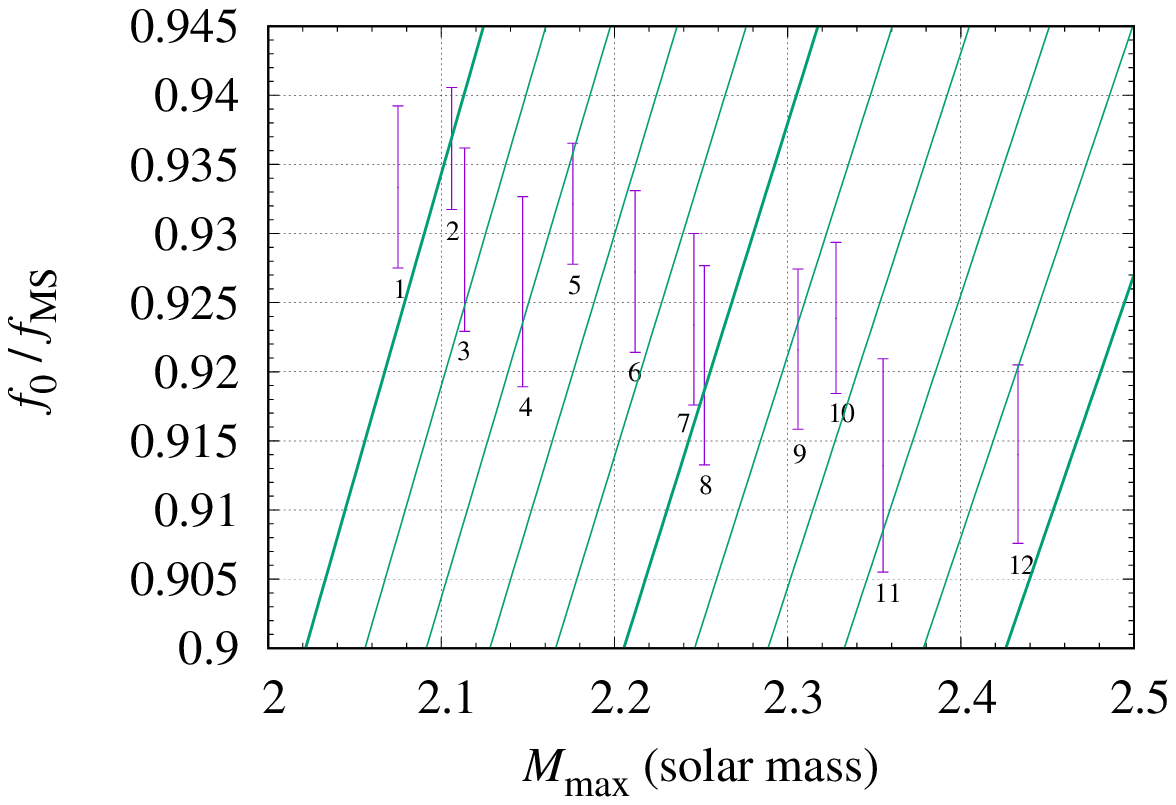}~~
\epsfxsize=3.4in \leavevmode \epsffile{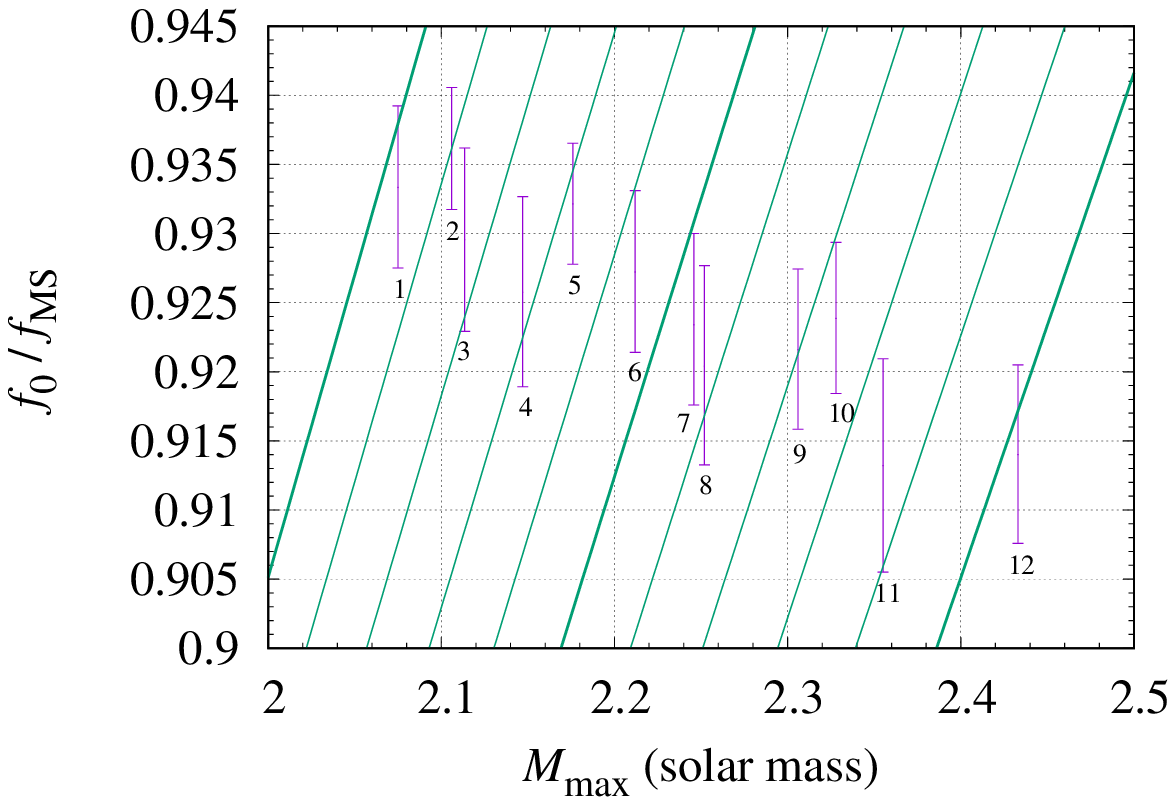}
\caption{Relation between $f_0/f_{\rm MS}$ and $M_{\rm max}$ for
  marginally-stable rigidly-rotating neutron stars with several
  equations of state listed in Table~\ref{table:EOS}. The lower and
  upper ends of each uncertainty range correspond to $J=0$ and
  $J=J_{\rm max}$, respectively. The numbers attached in each data
  denote the model equations of state.  The tilted lines show the
  relation of Eq.~(\ref{frrel}) for $f_r=1.20$, 1.18, 1.16, $\cdots$,
  1.04, 1.02, and 1.00 (from left to right) with $(M_{\rm out}+M_{\rm
    eje})/f_{\rm MS}=0.04M_\odot$ (left panel) and $0.08M_\odot$
  (right panel) and with $M=2.74M_\odot$.  The thick lines are for
  $f_r=1.20$, 1.10, and 1.00.
\label{fig2}}
\end{figure*}


In the analysis of Sec.~\ref{sec2}, we have several unknown
parameters; $f_0$, $f_{\rm MS}$, $E_{\rm GW,i}$, $E_{\rm GW,p}$,
$E_\nu$, $f$, $\Omega$, $M_{\rm MNS}$, $M_{\rm out}$, $M_{\rm eje}$,
$R_{\rm MNS}$, $R_{\rm out}$, and $J_0$. Among them, $f_0$, $f_{\rm
  MS}$, $\Omega$, $M_{\rm MNS}$, and $R_{\rm MNS}$ for a given
equation of state are calculated (at least with good approximation) by
constructing equilibrium states of non-rotating and rotating neutron
stars. Also, $E_{\rm GW,i}$, $f$, $M_{\rm out}$, $M_{\rm eje}$, and
$R_{\rm out}$ are approximately obtained with the help of
numerical-relativity simulations as already mentioned in
Sec.~\ref{sec2}. $J_0$ is also obtained accurately with the help of
numerical relativity.  The dependence of $J_0$ on total binary mass $m_0$, 
symmetric mass ratio
(hereafter referred to as $\eta\,(\leq 1/4)$), and $R_{1.35}$ is
determined by the results by numerical relativity
simulations~\cite{Kiuchi19} as
\beq 
J_0 \approx G c^{-1} m_0^2 \eta \left[ a_1 - a_2 \delta\eta +a_3
  \bar R_{1.35}^3 \left(1+a_4\delta\eta\right)\right],
\label{eq3.1}
\eeq
where $\bar R_{1.35}$ denotes $R_{1.35}$ in units of 10\,km,
$\delta\eta=\eta-1/4 (\leq 0)$, $a_1 \approx 3.32$, $a_2 \approx 31$,
$a_3 \approx 0.137$, and $a_4 \approx 27$. We note that $J_0$
increases with the increase of $R_{1.35}$ and with the decrease of
$\eta$ because the merger occurs at a more distant orbit for larger
stellar radii and for more asymmetric binaries.

Equation~(\ref{eq3.1}) shows that $J_0$ is in the range between
$\approx 5.8\times 10^{49}$\,erg\,s and $\approx 6.3\times
10^{49}$\,erg\,s for neutron stars with $R_{1.35}=10.5$--14\,km, total
mass $m_0 \approx 2.74M_\odot$, and $\eta=0.244$--0.250 (i.e., mass
ratio 0.73--1.00).  The value of $J_0$ is by $\sim 1 \times
10^{49}\,{\rm erg\,s}$ larger than the maximum angular momentum of
rigidly rotating neutron stars (see Fig.~\ref{fig1}) and $2T/\Omega$
for $\Omega \approx \Omega_K$~(see Eq.~(\ref{eq1.32})), and this fact
suggests that the remnant neutron star would initially have a
differentially rotating state with the maximum angular velocity 
slightly larger than $\Omega_K$, as has been found in many
numerical-relativity simulations since Ref.~\cite{shibata00}.


In contrast to $J_0$ and $E_{\rm GW,i}$, $E_{\rm GW,p}$ and
$E_\nu$ are not determined into a narrow range although $E_{\rm
  GW,p}+E_\nu$ is constrained by Eq.~(\ref{egwenu}) for given values
of $f_0/f_{\rm MS}$, $E_{\rm GW,i}$, and $(M_{\rm out}+M_{\rm
  eje})/f_{\rm MS}$ fairly well. We note that $E_\nu$ is associated
with the lifetime of the remnant neutron star of GW170817, $\tau$, as
$E_\nu \approx L_\nu \tau$. Here, the neutrino luminosity is $L_\nu
\agt 10^{53}$\,erg/s~\cite{Fujiba2019} but $\tau$ is not very
clear. Thus, we consider $E_{\rm GW,p}$ and $E_\nu$ as values to be
determined in the present analysis.

\begin{table*}[t]
 \caption{Key quantities for rigidly rotating neutron stars at the
   maximum mass along the marginally stable sequences for selected
   equations of state listed in Table~\ref{table:EOS}.  $M_{\rm
     MS,R}$: gravitational mass, $M_{*\rm MS,R}$: baryon rest mass,
   $J_{\rm MS,R}$: angular momentum, $\Omega_{\rm MS,R}$: angular
   velocity, $T_{\rm MS,R}$: rotational kinetic energy, and $R_{\rm
     MS,R}$: circumferential radius at the equatorial surface.  }
 \begin{tabular}{cccccccc} \hline
~~Model~~ & $M_{\rm MS,R}\,(M_\odot)$ & $M_{*\rm MS,R}/M_{\rm MS,R}$ 
& $M_{\rm MS,R}/M_{\rm max}$ 
& $J_{\rm MS,R}~(10^{49}\,{\rm erg\,s})$ & $\Omega_{\rm MS,R}$ ($10^4$\,rad/s)
& $T_{\rm MS,R}$\,($10^{53}$\,erg) & $R_{\rm MS,R}$\,(km) \\
 \hline \hline
EOS-1 & 2.497 & 1.185 & 1.203 & 3.875 & 1.160 & 2.247 & 13.04 \\
EOS-2 & 2.474 & 1.161 & 1.175 & 3.567 & 1.055 & 1.880 & 13.74 \\
EOS-3 & 2.570 & 1.193 & 1.216 & 4.183 & 1.155 & 2.415 & 13.15  \\ 
EOS-4 & 2.631 & 1.203 & 1.225 & 4.492 & 1.192 & 2.677 & 13.11  \\ 
EOS-5 & 2.574 & 1.167 & 1.183 & 3.919 & 1.028 & 2.014 & 14.19 \\
EOS-6 & 2.649 & 1.182 & 1.198 & 4.315 & 1.077 & 2.323 & 13.87 \\
EOS-7 & 2.707 & 1.188 & 1.205 & 4.580 & 1.090 & 2.496 & 13.93  \\ 
EOS-8 & 2.757 & 1.203 & 1.224 & 4.932 & 1.143 & 2.819 & 13.73 \\
EOS-9 & 2.770 & 1.185 & 1.208 & 4.756 & 1.049 & 2.494 & 14.33 \\
EOS-10& 2.774 & 1.176 & 1.191 & 4.668 & 1.007 & 2.349 & 14.79 \\
EOS-11& 2.907 & 1.209 & 1.234 & 5.548 & 1.099 & 3.049 & 14.26  \\ 
EOS-12& 2.945 & 1.195 & 1.210 & 5.504 & 1.039 & 2.858 & 14.77 \\
\hline
 \end{tabular}
 \label{table:EOS2}
\end{table*}

In the following, we constrain the value of $M_{\rm max}$ by analysing
rigidly rotating neutron stars at marginally stable states (at turning
points) for several equations of state.  For computing rigidly
rotating neutron stars in equilibrium, we employ a piecewise
polytropic equation of state, for which the details are described in
Appendix A. The selected equations of state used in this section are
listed in Table~\ref{table:EOS}. We selected many equations of state
rather randomly for each range of $M_{\rm max}$.  For the analysis, we
only employ the equations of state with which $2M_\odot$ neutron
stars~\cite{two} are reproduced, $\Lambda_{1.35} \leq 1000$, and the
sound speed is always smaller than the speed of light for stable
neutron stars.  We ignore the thermal effect of the remnant neutron
star for constructing equilibrium rotating neutron stars because it is
a minor contribution to the neutron-star properties if we consider its
age to be of order 0.1--1\,s~\cite{Kaplan14}.

Figure~\ref{fig1} displays several key quantities as functions of the
angular momentum, $J$, along the sequences of the marginally stable
neutron stars (cf. the dashed curve of Fig.~\ref{fig0}). Here, the
neutron stars with $J=0$ denote the marginally stable state for the
spherical neutron stars of mass $M_{\rm max}$, and at $J=J_{\rm max}$,
the neutron stars are at the mass-shedding limit; the angular velocity
at the equatorial surface is equal to the Keplerian one.
Table~\ref{table:EOS2} also lists several quantities of rigidly
rotating neutron star at a turning point and mass shedding limit.

Figure~\ref{fig2} displays the relation between $f_0/f_{\rm MS}$ and
$M_{\rm max}$ for marginally-stable rigidly-rotating neutron stars
with several piecewise polytropic equations of state.  Here, $f_0$ is
calculated for binaries of mass ($1.35_\odot, 1.40M_\odot$) and
($1.20_\odot, 1.55M_\odot$), but the values for two cases are
different only by 0.2--0.3\% (see Fig.~\ref{figa1}), and thus, the
difference in $f_0$ is not very important in the following.  The
uncertainty in Fig.~\ref{fig2} reflects the variation of $f_{\rm MS}$
for the different angular momentum of the marginally stable rotating
neutron stars: As found from Fig.~\ref{fig1}, $f_0/f_{\rm MS}$ is
smallest for $J=0$ and largest for $J=J_{\rm max}$.  The tilted lines
show the relation of Eq.~(\ref{frrel}) for $f_r=1.20-0.02 \times i$ up
to 1.00 (from left to right) with $(M_{\rm out}+M_{\rm eje})/f_{\rm
  MS}=0.04M_\odot$ and $0.08M_\odot$, while $M$ is fixed to be
$2.74M_\odot$. We note that for a given value of $M_{\rm max}$, the
value of $f_0/f_{\rm MS}$ is in general larger for the larger values
of $\Lambda_{1.35}$, and hence, the upper bound is determined by the
constraint for $\Lambda_{1.35}$ (see Appendix A).

For rigidly rotating neutron stars, the upper limit of $f_r$ is
approximately 1.2 as found from Fig.~\ref{fig1} (see also
Table~\ref{table:EOS2}). Thus, the left region of the line of
$f_r=1.2$ in Fig.~\ref{fig2} is prohibited in our model. That is, for
the equations of state with small values of \ms{$M_{\rm max}=
2.0$--$2.1M_\odot$}, the remnant neutron star cannot achieve the
rigidly rotating state at the onset of collapse (we show the examples
in Sec.~\ref{sec3c}). For such equations of state, the remnant neutron
star would collapse before the rigidly rotating state is achieved,
i.e., its lifetime is shorter than the angular momentum transport
timescale in the remnant neutron star. It is natural to consider that
the lifetime of the remnant neutron star with such equations of state
is fairly short $\ll 1$\,s.

For the right region of the line of $f_r=1$, on the other hand, the
collapse cannot occur, because the mass of the remnant neutron star is
smaller than $M_{\rm max}$.  Thus, for such equations of state, a
stable neutron star should be the final outcome. As mentioned in the
first paragraph of Sec.~\ref{sec2}, if such a stable remnant is formed
in GW170817 and has large rotational kinetic energy $\agt
10^{52}$\,erg, an energy injection to ejecta through electromagnetic
radiation like the magnetic dipole radiation would occur and be
inconsistent with the observational results for the electromagnetic
counterparts of GW170817~\cite{MM2017,shibata17}. However, if the
rotational kinetic energy is dissipated or removed in a short
timescale ($\alt 100$\,s), e.g., by gravitational radiation and
neutrino emission, we may accept the formation of a stable neutron
star~\cite{Li2018}.

Figure~\ref{fig2} shows an important fact as follows: If the maximum
mass of spherical neutron stars in nature is relatively small $\alt 
2.1M_\odot$, the collapse would occur for the remnant neutron star in a
rapidly rotating state with $f_r \approx 1.2$. On the other hand, if
the maximum mass of spherical neutron stars is relatively large $\agt
2.3M_\odot$, the collapse would not occur for rapidly rotating remnant
neutron star but for $f_r \alt 1.05$. Then the next issue is whether
we can find a self-consistent solution for such collapses, because
$f_r$ is determined by the angular momentum dissipation process in the
post-merger stage. In the following, we show several realistic
scenarios for this process in some of equations of state that we
select.

\subsection{Method}\label{sec3b}

For each equation of state, we try to find a solution that satisfies
the conservation relations of energy and angular momentum
self-consistently. \ez{We will have a brief summary of our model by analyzing 
the degree of freedoms of all unknown variables here.
There are in total 14 variables in the conservation law
of rest mass, energy and angular momentum (cf. Eq.~(\ref{eq1.10}), ~(\ref{eq1.1})
and (\ref{eq1.4})). Nevertheless, $M$ is determined by observation and $M_*$ is
related to $M$ by factor $f_0$ which could be determined once a model equation of state is
assumed. Similarly, according to the marginally stable solution results, $M_{f}$ and $J_{f}$
are fixed once $M_{*f}$ is fixed and an equation of state is given. Hence, there is only
one degree of freedom among $M_{f}$, $M_{*f}$ and $J_{f}$. $J_0$ is given 
by Eq.~(\ref{eq3.1}). There are 8 other variables left: $E_\nu$ and $J_\nu$ are related 
by Eq.~(\ref{Jnu}),
$E_{\rm GW,p}$ and $J_{\rm GW,p}$ are related
by Eq.~(\ref{Jgwp}), $M_{\rm eje}+M_{\rm out}$ and $J_{\rm eje}+J_{\rm out}$ are 
related by Eq.~(\ref{Jeje}) and (\ref{Jout}). As a consequence, there are only 3 degrees
of freedom in the remaining 8 variables. In total, there are 4 degrees of freedom and we 
have 3 equations (laws of conservation). Once we have input one parameter,
such as $M_{\rm eje}+M_{\rm out}$,  and a model equation of state
is given, all the other quantities related to the properties of the star
at the moment of collapse, the total energy of neutrino emission and gravitational wave in the
post-merger phase could be solved (cf. Table.~\ref{table:EOS3}).
With this analysis we can tell whether the assumed equation of state model is
consistent with the observation, for instance, the obtained values of $E_\nu$ and $E_{\rm GW,p}$ have to
be positive and within a plausible range.}

In this analysis, we employ $M_f$, $R_{\rm MS}$
(circumferential radius), and $\Omega_{\rm MS}$ (angular velocity) for
rigidly rotating neutron stars at turning points as $M_{\rm MNS}$,
$R_{\rm MNS}$, and $\Omega_{\rm MNS}$, respectively.  $f_{\rm MS}$ is
also found from a solution of rigidly rotating neutron stars at
turning points.  $f_0$ is calculated from the solutions of spherical
neutron stars of mass 1.35-$1.40M_\odot$ or 1.20-$1.55M_\odot$.
$E_{\rm GW,i}$ is set to be $(0.040 \pm 0.005)M_\odot c^2$.  $J_0$ is
determined by Eq.~(\ref{eq3.1}) for a given value of $R_{1.35}$ and
$\eta$.  Here the variation in $\eta$ by $\pm 0.003$ systematically
changes the values of $E_{\rm GW,p}$ and $E_\nu$ only by at most $\pm
0.007M_\odot$ and $\mp 0.007M_\odot$.  Hence, we only show the results
with $\eta=0.247$.  Since the dependence of $J_{\rm out}+J_{\rm eje}$
on $R_{\rm out}$ is not very strong (compared with the dependence on
$M_{\rm out}+M_{\rm eje}$), we fix $R_{\rm out}$ to be 140\,km.  On
the other hand, $M_{\rm out}+M_{\rm eje}$ has a strong effect on the
solution. Thus, we vary it in a wide range (e.g., see
Fig.~\ref{fig3}).

With these preparations, Eq.~(\ref{frrel}) can be considered as a
relation between $f_r$ and $f_{\rm MS}$. Here, in Eq.~(\ref{frrel}),
$M$ is given ($2.74M_\odot$), $f_0$ is determined for a given equation
of state and each mass of binary, and $M_{\rm out}+M_{\rm eje}$ is an
input parameter.  An equilibrium sequence of rigidly rotating neutron
stars along the turning points for a given equation of state also
gives another monotonic relation between these two variables as found
in Fig.~\ref{fig1}. Thus, we first determine $f_r$ and $f_{\rm MS}$ by
solving a simultaneous equation composed of two independent relations
and specify a model for the rigidly rotating neutron star at a turning
point. (We note that for some equations of state, e.g., EOS-12, the
solution does not exist.) We can then obtain $J_f$, $M_f$,
$\Omega_{\rm MS}$, and $R_{\rm MS}$ for this model using the monotonic
relations of $f_{\rm MS}(J_f)$, $M_f(J_f)$, $\Omega_{\rm MS}(J_f)$,
and $R_{\rm MS}(J_f)$: cf. Fig.~\ref{fig1}.

We can subsequently determine $E_{\rm GW,p}+E_\nu$ and $J_{\rm
  GW,p}+J_{\nu}$ from Eqs.~(\ref{egwenu}) and (\ref{eq1.4}).  For
Eq.~(\ref{eq1.4}), we employ $J_0$, $M_{\rm MNS}(=M_f)$, $R_{\rm
  out}$, $R_{\rm MNS}(=R_{\rm MS})$, $f$, and $\Omega(=\Omega_{\rm
  MS})$ for each equation of state.  Then these two relations, $E_{\rm
  GW,p}+E_\nu=$const and $J_{\rm GW,p}+J_\nu=$const, constitute a
simultaneous equation for $E_{\rm GW,p}$ and $E_\nu$ because we have
already given the values of $R_{\rm MNS}$, $\Omega$, and $f$, which
are necessary to relate $J_{\rm GW,p}$ and $J_\nu$ to $E_{\rm GW,p}$
and $E_\nu$, respectively. Thus, these two quantities are immediately
determined, if the solution exists. (Again we note that for some
equations of state, a physical solution does not exist: see below.)

\subsection{Results}\label{sec3c}

\begin{table}[]
 \caption{Predicted states of rigidly rotating neutron stars at the
   onset of collapse for the GW170817 event with several equations of
   state.  The given values of $M_{\rm out}+M_{\rm eje}$ are
   $0.048M_\odot$ (upper), $0.096M_\odot$ (middle), and $0.150M_\odot$
   (lower), respectively.  $M_f$ and $J_f$ are shown in units of
   $M_\odot$ and $10^{49}$\,erg\,s, respectively.  The units of
   $E_{\rm GW,p}$ and $E_\nu$ are $M_\odot c^2$.  ``---'' means that
   no solution of $f_r$ and $f_0/f_{\rm MS}$ exists for the
   corresponding equations of state.  Associated with the
   uncertainties in $J_0$ and $E_{\rm GW,i}$ by $\pm 0.1 \times
   10^{49}$\,erg\,s, and $0.005M_\odot c^2$, respectively, an
   uncertainty, typically, of $\pm 0.007M_\odot c^2$ and $\mp
   0.012M_\odot c^2$, exists in $E_{\rm GW,p}$ and $E_\nu$,
   respectively.  $^{*}$ shows that only solutions with negative
   values of $E_\nu$ are obtained. For EOS-12, no solution for $f_r$
   and $f_0/f_{\rm MS}$ exists for $M_{\rm out}+M_{\rm eje} \agt
   0.058M_\odot$ and no solution with the positive value of $E_\nu$
   exists for $M_{\rm out}+M_{\rm eje} \leq 0.15M_\odot$, so that we do
   not describe the results.}
 \begin{tabular}{ccccccc} \hline
~~Model~~ & $f_0/f_{\rm MS}$ & $f_r$ 
& ~~$M_f$~~ & ~~$J_f$~~ & ~$E_{\rm GW,p}$~ & ~$E_\nu$~  \\  \hline \hline
EOS-1 & --- & --- & --- & --- & ---  & ---  \\
EOS-2 & --- & --- & --- & --- & ---  & ---  \\
EOS-3 & 0.935 & 1.194 & 2.52 & 3.89 & 0.076  & 0.053 \\
EOS-4 & 0.931 & 1.169 & 2.51 & 3.71 & 0.091  & 0.049 \\
EOS-5 & 0.936 & 1.160 & 2.52 & 3.59 & 0.082  & 0.047 \\
EOS-6 & 0.930 & 1.134 & 2.51 & 3.35 & 0.102  & 0.043 \\
EOS-7 & 0.926 & 1.111 & 2.50 & 3.11 & 0.118  & 0.038 \\
EOS-8 & 0.921 & 1.103 & 2.48 & 3.01 & 0.130  & 0.037 \\
EOS-9 & 0.922 & 1.077 & 2.49 & 2.65 & 0.130  & 0.037 \\
EOS-10& 0.923 & 1.069 & 2.49 & 2.51 & 0.130  & 0.034 \\
EOS-11& 0.910 & 1.042 & 2.45 & 1.98 & 0.194  & 0.004 \\
\hline
EOS-1  & 0.939 & 1.201 & 2.49 & 3.84 & 0.034  & 0.078 \\
EOS-2  & --- & --- & --- & --- & ---  & ---  \\
EOS-3  & 0.935 & 1.174 & 2.48 & 3.62 & 0.051  & 0.072 \\
EOS-4  & 0.930 & 1.150 & 2.47 & 3.43 & 0.067  & 0.068 \\
EOS-5  & 0.935 & 1.140 & 2.48 & 3.30 & 0.064  & 0.059 \\
EOS-6  & 0.929 & 1.114 & 2.46 & 3.04 & 0.083  & 0.057 \\
EOS-7  & 0.924 & 1.092 & 2.45 & 2.78 & 0.100  & 0.051 \\
EOS-8  & 0.920 & 1.084 & 2.44 & 2.67 & 0.112  & 0.051 \\
EOS-9  & 0.920 & 1.058 & 2.44 & 2.26 & 0.118  & 0.045 \\
EOS-10 & 0.921 & 1.050 & 2.44 & 2.09 & 0.120  & 0.040 \\
EOS-11 & 0.908 & 1.023 & 2.41 & 1.44 & 0.188  & 0.006 \\
\hline
EOS-1  & 0.938 & 1.178 & 2.44 & 3.54 & 0.008  & 0.098 \\
EOS-2  & 0.940 & 1.162 & 2.45 & 3.39 & 0.024  & 0.079 \\
EOS-3  & 0.934 & 1.151 & 2.43 & 3.31 & 0.025  & 0.092 \\
EOS-4  & 0.929 & 1.127 & 2.42 & 3.10 & 0.041  & 0.088 \\
EOS-5  & 0.935 & 1.118 & 2.43 & 2.96 & 0.046  & 0.071 \\
EOS-6  & 0.928 & 1.092 & 2.42 & 2.68 & 0.064  & 0.070 \\
EOS-7  & 0.923 & 1.070 & 2.40 & 2.38 & 0.083  & 0.063 \\
EOS-8  & 0.918 & 1.062 & 2.39 & 2.25 & 0.094  & 0.064 \\
EOS-9  & 0.918 & 1.037 & 2.39 & 1.75 & 0.109  & 0.051 \\
EOS-10 & 0.920 & 1.028 & 2.39 & 1.54 & 0.114  & 0.043 \\
EOS-11$^{*}$ & --- & --- & --- & --- & ---  & --- \\
\hline\hline
 \end{tabular}
 \label{table:EOS3}
\end{table}

\begin{figure*}[t]
\epsfxsize=3.4in \leavevmode \epsffile{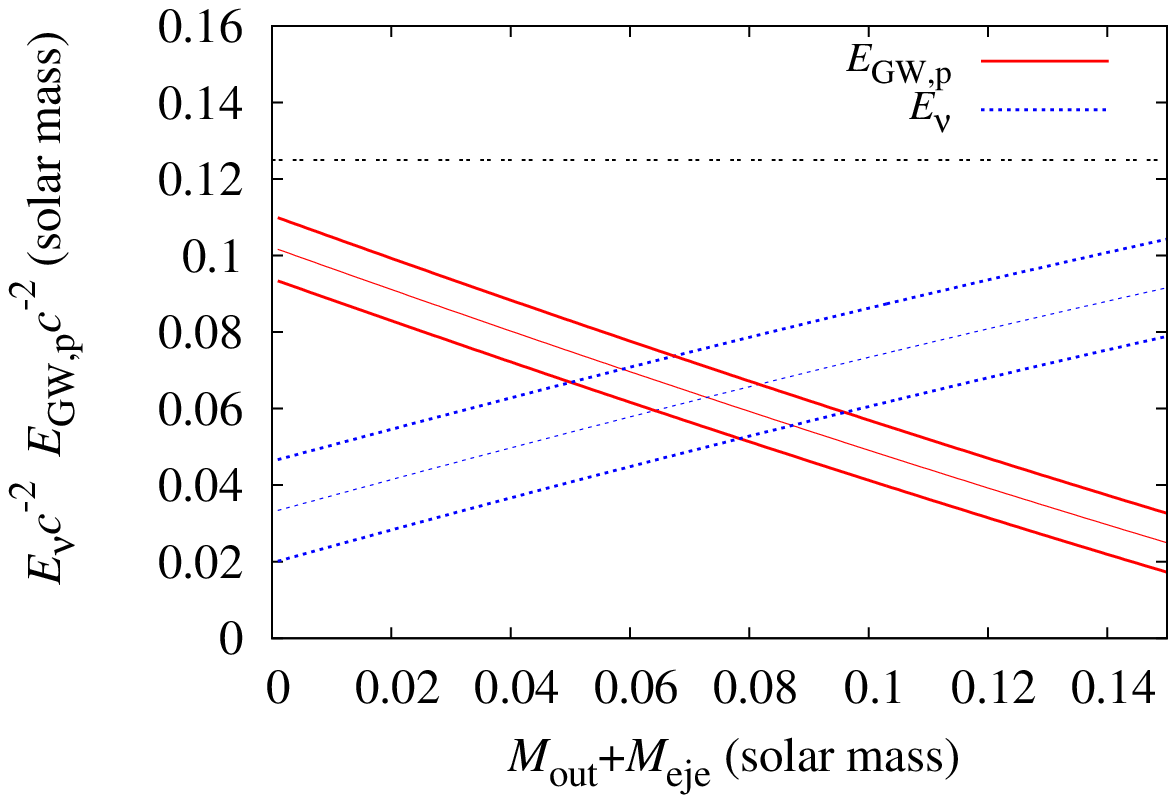}~~
\epsfxsize=3.4in \leavevmode \epsffile{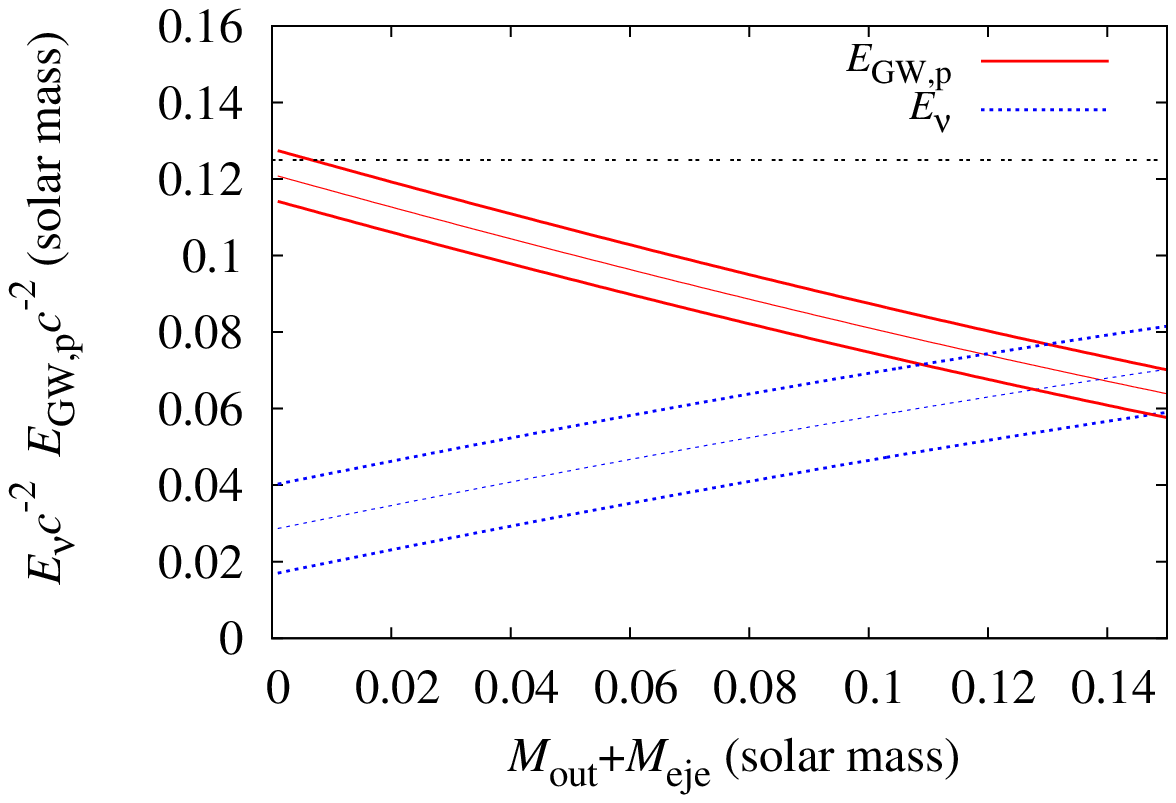} \\
\epsfxsize=3.4in \leavevmode \epsffile{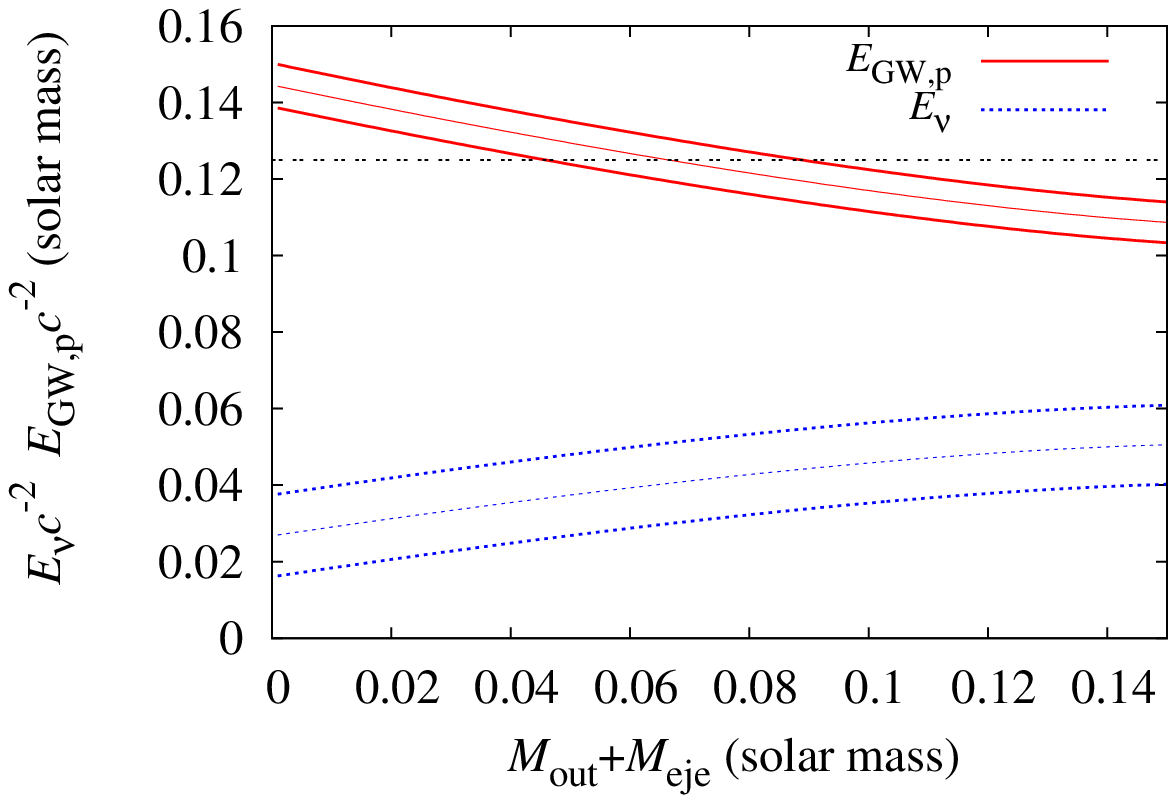}~~
\epsfxsize=3.4in \leavevmode \epsffile{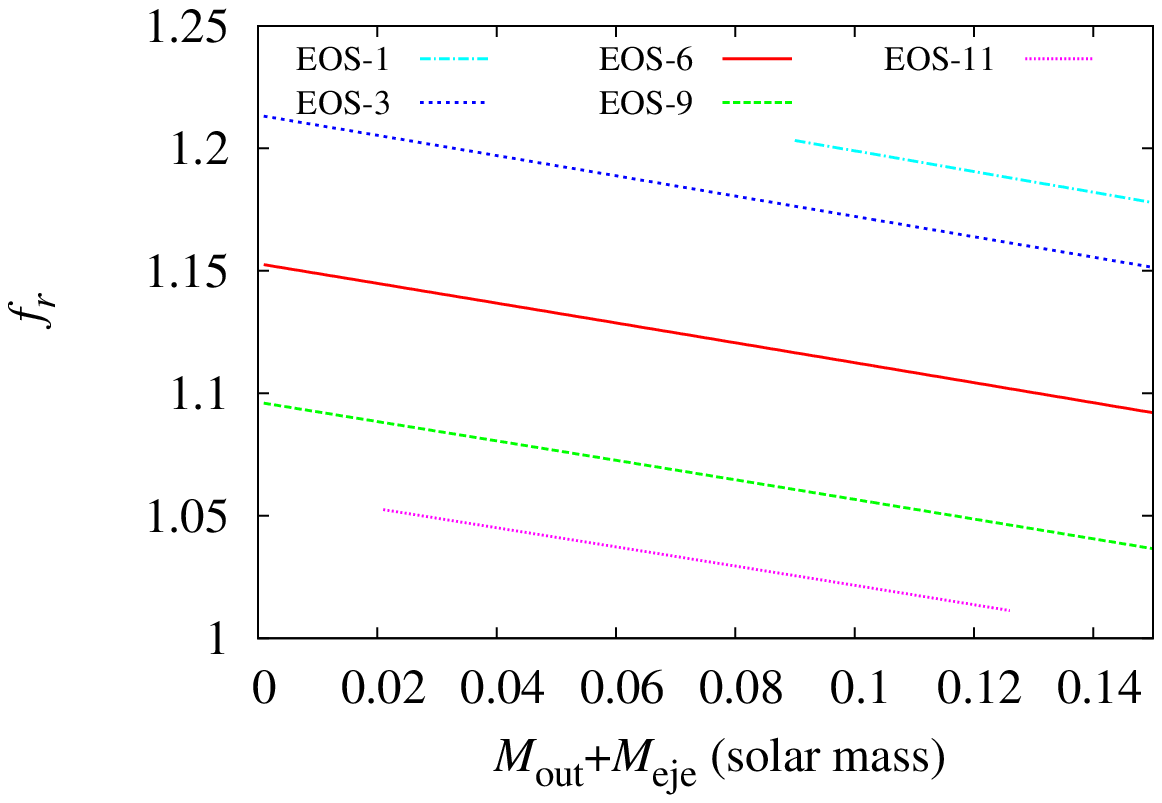}
\caption{ Top-left, top-right, and bottom-left panels display $E_{\rm
    GW,p}$ and $E_\nu$ as functions of $M_{\rm out}+M_{\rm eje}$ for
  EOS-3, 6, and 9, respectively, and for the $1.35$-$1.40M_\odot$
  case.  Associated with the uncertainties in $J_0$ and $E_{\rm
    GW,i}$, an uncertainty exists in $E_{\rm GW,p}$ and $E_\nu$. The
  three curves for each plot in the the top-left, top-right, and
  bottom-left panels denote the upper and lower bounds (thick curves)
  as well as the central value (thin curve) for $E_{\rm GW,p}$ and
  $E_\nu$. The horizontal dot-dot line shows $E_{\rm
    GW,p}=0.125M_\odot c^2$ (plausible upper limit of $E_{\rm
    GW,p}$~\cite{Zappa}).  Bottom-right panel shows $f_r$ as a
  function of $M_{\rm out}+M_{\rm eje}$ for EOS-1, 3, 6, 9, and 11.
  For EOS-1, the solution for $f_r$ does not exist for small value of
  $M_{\rm out}+M_{\rm eje}$.  For EOS-11, $E_\nu$ becomes negative for
  $M_{\rm out}+M_{\rm eje}\leq 0.02M_\odot$ and $\geq 0.127M_\odot$,
  and thus, we do not plot the solution for these ranges.
\label{fig3}}
\end{figure*}


Table~\ref{table:EOS3} shows the solutions that self-consistently
satisfy the conservation relations of energy and angular momentum for
each equation of state with the selected values of $M_{\rm out}+M_{\rm
  eje}$; $0.048M_\odot$, $0.096M_\odot$, and
$0.150M_\odot$. Figure~\ref{fig3} also shows representative results:
The top-left, top-right, and bottom-left panels display $E_{\rm GW,p}$
and $E_\nu$ as functions of $M_{\rm out}+M_{\rm eje}$ for EOS-3, 6,
and 9, respectively, and the bottom-right panel shows $f_r$ as a
function of $M_{\rm out}+M_{\rm eje}$ for EOS-1, 3, 6, 9, and 11.
Here, associated with the uncertainties in $J_0$ and $E_{\rm GW,i}$ by
$\pm 0.1 \times 10^{49}$\,erg\,s, and $0.005M_\odot c^2$,
respectively, an uncertainty, typically, of $\pm 0.007M_\odot c^2$ and
$\mp 0.012M_\odot c^2$, exists in $E_{\rm GW,p}$ and $E_\nu$,
respectively. The three curves for each plot in the the top-left,
top-right, and bottom-left panels of Fig.~\ref{fig3} denote the upper
and lower bounds as well as the central value for $E_{\rm GW,p}$ and
$E_\nu$.  In addition, the change of $f_0$ from the 1.35-$1.40M_\odot$
case to the 1.20-$1.55M_\odot$ case varies $E_{\rm GW,p}$ and $E_\nu$
typically by $-0.003M_\odot c^2$ for both quantities.

For EOS-12, no solution is found for given parameters. The reason for
this is that (i) for large values of $M_{\rm out} + M_{\rm eje}$, the
predicted final mass of the remnant neutron star, $M_f$, becomes
smaller than $M_{\rm max}$, and hence, no solution with $f_r \geq 1$
is present and (ii) for $M_{\rm out}+M_{\rm eje} \alt 0.057M_\odot$,
the value of $f_r$ is determined but physical (positive) values for
the set of ($E_{\rm GW,p}, E_\nu$) are not found. If this type of
equation of state with $M_{\rm max} \agt 2.4M_\odot$ would be the real
one, the final outcome should be a stable neutron star in the GW170817
event. However, this is not likely as we discuss in Sec.~\ref{sec3f}.


For EOS-1 and 2 for which $M_{\rm max}$ is rather small, $\alt
2.1M_\odot$, we often fail to find a solution. The reason for this is
that Eq.~(\ref{frrel}) can be satisfied only for a high value of $f_r
\sim 1.2$, and thus, for small values of $M_{\rm out}+M_{\rm eje} \alt
0.09M_\odot$, any solution cannot be present (cf.~Fig.~\ref{fig2} and
the bottom-right panel of Fig.~\ref{fig3}).  However, this fact does
not imply that these equations of state with $M_{\rm max} \alt
2.1M_\odot$ cannot be accepted.  It is still possible that our
assumption described in the first paragraph of Sec.~\ref{sec2} might
be inappropriate \ms{(i.e., the collapse of the remnant neutron star
  to a black hole might occur before a rigidly rotating state was
  reached)}.  If the equation of state with a low value of $M_{\rm
  max}$ would be the real one, the collapse to a black hole could
occur in a stage that the remnant neutron star is differentially
rotating for the GW170817 event. If so, the value of $f_r$ is larger
than $\agt 1.2$, and thus, the collapse would occur within the angular
momentum transport timescale of the remnant neutron star.


Table~\ref{table:EOS3} and Fig.~\ref{fig3} show that for EOS-3--7 for
which $M_{\rm max}=2.10$--$2.25M_\odot$, we find solutions with
plausible values of $E_{\rm GW,p}$ and $E_\nu$ as $0.03M_\odot c^2
\alt E_{\rm GW,p} \alt 0.11M_\odot c^2$ and $0.03M_\odot c^2 \alt
E_\nu \alt 0.10M_\odot c^2$ for plausible values of $M_{\rm
  out}+M_{\rm eje}$.  Also, as Fig.~\ref{fig3} illustrates for EOS-3
and 6, reasonable solutions exist for a wide range of $M_{\rm
  out}+M_{\rm eje}$ for this class of the equations of state.  Here,
the value of $E_\nu$ indicates that the predicted lifetime of the
remnant neutron star in these equations of state is of order 0.1\,s to
1\,s, which is also quite reasonable.  Thus, we conclude that we do not
have any reason to exclude equations of state with $2.10M_\odot \alt
M_{\rm max} \alt 2.25M_\odot$. It should be emphasized that for these
cases, the value of $f_r$ is not always close to $\sim 1.2$ but in a
wide range between 1.07 and 1.20 (see the bottom-right panel of
Fig.~\ref{fig3}). Thus, $f_r \approx 1.2$ does not always hold for
many candidate equations of state.

For EOS-9 and 10 for which $M_{\rm max} \approx
2.30M_\odot$--$2.33M_\odot$, we can also find solutions. However, for
these cases, a highly efficient energy dissipation by gravitational
radiation with $E_{\rm GW,p} \agt 0.11M_\odot$ is necessary in
particular for small values of $M_{\rm out}+M_{\rm eje}$ (see the
bottom-left panel of Fig.~\ref{fig3}).  As found from
Table~\ref{table:EOS2}, the maximum rotational kinetic energy of the
rigidly rotating neutron stars is $T_{\rm MS,R} \approx 2.5 \times
10^{53}\,{\rm erg} \approx 0.125M_\odot c^2$.  This implies that the
energy dissipated by gravitational radiation has to be comparable to
the rotational kinetic energy of the remnant neutron star. To know if
this class of equations of state is really viable, we have to perform a
numerical-relativity simulation to check whether such efficient
gravitational radiation is possible or not. However, this is beyond
the scope of this paper and is left for our future study.  The
bottom-right panel of Fig.~\ref{fig3} illustrates that for this class
of equations of state, the value of $f_r$ has to be small ($\leq
1.1$). That is, an efficient angular momentum dissipation is 
supposed. 

For EOS-8 of $M_{\rm max} \approx 2.25M_\odot$, we also find solutions
with high values of $E_{\rm GW,p}$.  However, for this case, with 
high values of $M_{\rm out}+M_{\rm eje}$, the required value of $E_{\rm
  GW,p}$ can be reduced.  Also, the maximum rotational kinetic energy
of the rigidly rotating neutron stars for this equation of state is
relatively high, $T_{\rm MS,R} \approx 2.8 \times 10^{53}\,{\rm erg}
\approx 0.14M_\odot c^2$. Thus, the restriction for this equation of
state is not as strong as for EOS-9 and 10.


For EOS-11 for which $M_{\rm max} \approx 2.35M_\odot$, the
required energy dissipated by gravitational radiation, $E_{\rm GW,p}$,
far exceeds $0.125M_\odot c^2$ that is a plausible maximum
value~\cite{Zappa}. Moreover, for this case, $E_{\rm GW,p}$ required
exceeds the maximum rotational kinetic energy of the rigidly rotating
neutron stars, $T_{\rm MS,R}$. The reason why the required value of
$E_{\rm GW,p}$ is very large is that for this case the value of
$f_r$ is quite small, $\leq 1.05$, and thus, a large fraction of the
angular momentum dissipation, $\sim 4 \times 10^{49}$\,erg\,s, is
necessary to reach a marginally stable state. However, for such
significant angular momentum dissipation, unrealistically large
dissipation by gravitational radiation is necessary.  Therefore it is
reasonable to exclude these equations of state.

To conclude, it is easy to find model equations of state with $M_{\rm
  max} \leq 2.25M_\odot$ that satisfy the conservation laws of energy
and angular momentum self-consistently. Also it is not impossible to
find model equations of state with $M_{\rm max} \alt 2.3M_\odot$ that
satisfy the required laws. For these cases, $f_r$ is not always
$\approx 1.2$.  By contrast, it would not be easy to find an equation
of state with $M_{\rm max} \geq 2.35M_\odot$ that satisfy the required
laws.

\subsection{Stable neutron star formation: not likely}\label{sec3f}

For $M_{\rm max} \agt 2.4M_\odot$, a stable neutron star could be the
final outcome (e.g., EOS-12) as already mentioned above. For this case
to be viable, its angular momentum (and rotational kinetic energy) has
to be sufficiently small, since the observational results for the
electromagnetic counterparts of GW170817 do not show the evidence for
the energy injection to ejecta from strong electromagnetic radiation
like the magnetic dipole radiation associated with the rotational
kinetic energy of the remnant neutron star~\cite{MM2017}. If we
require that the resulting rotational kinetic energy of the stable
neutron star at its age of $\sim 100$\,s is smaller than
$10^{52}$\,erg (i.e., by one order of magnitude smaller than the
rest-mass energy of ejecta of mass $\sim 0.05M_\odot$), we need $J_f <
0.5 \times 10^{49}$\,erg\,s (see Fig.~\ref{fig1}).  Because $J_0
\approx 6.0 \times 10^{49}$\,erg\,s for stiff equations of state like
EOS-12, we obtain a constraint as
\beqn
J_{\rm GW,p}+J_\nu+J_{\rm out}+J_{\rm eje} \agt 5.5 \times 10^{49}\,{\rm erg\,s}
\approx 0.9J_0.\nonumber \\ 
\label{eq3b5}
\eeqn
Thus, it is necessary for the remnant neutron star to relax to 
a fairly slow rotation state close to a spherical star. 

Figure~\ref{figa1} shows $f_0/f_{\rm MS} \geq 0.90$ ($f_{\rm MS} \sim
1.2$) for spherical neutron stars of $M_{\rm max} \approx 2.4M_\odot$.  Then
we also obtain the following conditions from Eq.~(\ref{egwenu}) and
$E_{\rm GW,i}=(0.040 \pm 0.005)M_\odot$:
\beqn
&&(E_{\rm GW,p}+E_\nu)c^{-2}+(M_{\rm out}+M_{\rm eje})
\left(1-{1 \over f_{\rm MS}}\right) \nonumber \\
&\leq& (0.234 \pm 0.005) M_\odot. \label{eq3a8}
\eeqn
Here, we supposed that the remnant neutron star would be located along
a stable branch near the marginally stable sequence, and hence, we
employed the equations derived in Sec.~\ref{sec2}.

In the hypothesis of this subsection, the remnant is long-lived and it
is natural to suppose that $E_\nu \sim 0.1M_\odot c^2 \approx 2\times
10^{53}$\,erg or more~\cite{Fujiba2019}. Note that it is often
mentioned that the total energy dissipated by the neutrino emission
from a protoneutron star formed in each supernova would be $\sim
(2$--$3) \times 10^{53}$\,erg (e.g., Ref.~\cite{Suwa19}). Here, the
value is larger for larger mass of the protoneutron stars.  The
remnant neutron star of binary neutron star mergers is more massive
and hotter than the protoneutron star, and hence, it is natural to
consider $E_\nu \agt 3\times 10^{53}$\,erg. In the following, we
conservatively assume that $E_\nu=3 \times 10^{53}\,{\rm erg} \approx
0.15M_\odot c^2$. For this case, we can estimate as $J_\nu \sim
0.5\times 10^{49}$\,erg\,s using Eq.~(\ref{Jnu}).

We also suppose $M_{\rm out} \ll M_{\rm eje}$ for $\tau \agt 100$\,s in
the following because the torus matter would accrete onto the neutron
star or be ejected from the system by viscous angular momentum
transport and/or propeller effect~\cite{PO11}.

Equation~(\ref{eq3a8}) with $M_{\rm out}=0$ gives a constraint as 
$(E_{\rm GW,p}+E_\nu)c^{-2} + M_{\rm eje}(1-1/f_{\rm MS}) \leq
0.24M_\odot$.  For $E_\nu=0.15M_\odot c^2$, we obtain
\beq
E_{\rm GW,p}c^{-2} + M_{\rm eje}\left(1-{1 \over f_{\rm MS}}\right) \leq 0.09M_\odot. 
\eeq

Now we consider two extreme cases (assume $f_{\rm MS}=1.2$): $E_{\rm
  GW,p}=0.085M_\odot c^2$ and $M_{\rm eje} = 0.03M_\odot$ (minimum
ejecta mass required for the GW170817 event), and $E_{\rm
  GW,p}=0.055M$ and $M_{\rm eje} = 0.15M_\odot$. Here, for the
GW170817 event, the value of $M_{\rm eje}$ would be smaller than
$0.15M_\odot$.

For $E_{\rm GW,p} = 0.085M_\odot c^2$ and $f=2.5$\,kHz (which would be
the possible lowest value), $J_{\rm GW,p} \approx 1.9\times
10^{49}$\,erg\,s.  For $M_{\rm eje}=0.03M_\odot$, the ejection of the
angular momentum would be $\approx 0.5 \times 10^{49}$\,erg\,s for
$R_{\rm eje} \approx 200$\,km (cf.~Eq.~(\ref{Jeje})).  Since $J_\nu
\approx 0.5\times 10^{49}$\,erg\,s, the constraint of Eq.~(\ref{eq3b5})
cannot be satisfied in this model at all.

For $M_{\rm eje}=0.15M_\odot$, the ejection of the angular momentum
would be $\approx 2.5 \times 10^{49}$\,erg\,s for $R_{\rm eje} \approx
200$\,km (e.g., Eq.~(\ref{Jeje})).  With $E_{\rm GW,p}=0.055M_\odot c^2$ and
$f=2.5$\,Hz, $J_{\rm GW,p} \approx 1.3\times 10^{49}$\,erg\,s. 
For $J_\nu \approx 0.5\times 10^{49}$\,erg\,s, it is found that 
the constraint of Eq.~(\ref{eq3b5}) cannot be also satisfied in this model.

As found from the above analysis, for a larger value of $M_{\rm eje}$,
$J_{\rm GW,p}+J_\nu+J_{\rm eje}$ increases. However, for the GW170817
event, $M_{\rm eje}$ is not very likely to be larger than
$0.15M_\odot$.  If $f$ is smaller than 2.5\,kHz, $J_{\rm GW,p}$ would
be larger. However, a number of numerical-relativity simulations have
shown $f \geq 2$\,kHz~\cite{Bauswein,Hotoke13,NR}; $J_{\rm GW,p}$
could be increased only by 20\%.  Thus, we conclude that it is quite
difficult to find a mechanism of the angular momentum dissipation by
$\approx 0.9J_0$ in a short timescale of $\sim 100$\,s.

Since the values of $E_\nu$, $R_{\rm eje}$, $M_{\rm eje}$, and $f$
have uncertainty, it is not possible to fully exclude the possibility
of forming a stable neutron star. In particular, in case that $E_\nu$
could be much smaller than $3 \times 10^{53}$\,erg, the angular
momentum of the remnant neutron star could be smaller than $0.1J_0$,
e.g., by setting $E_{\rm GW,p}=0.125M_\odot c^2$, $E_\nu=0.09M_\odot
c^2$, and $M_{\rm eje}=0.15M_\odot$ with $f=2.5$\,kHz and $R_{\rm
  eje}=200$\,km.  However, we need a (unphysical) fine tuning, and
hence, such a possibility would not be very likely.

\section{Summary}\label{sec4}

We study the constraint on the maximum mass of cold spherical neutron
stars coming from the observational results of GW170817 more strictly
than our previous study. We develop a framework which employs not only
energy conservation law but also angular momentum conservation one, as
well as solid results of latest numerical-relativity simulations and
of neutron stars in equilibrium.

In this framework, we postulate that a massive neutron star was formed
as a remnant after the merger in the GW170817 event, and the collapse
occurred after the remnant neutron star relaxes to a rigidly rotating
state. Thus, we construct several rigidly rotating neutron stars in
equilibrium as models of the remnant neutron stars at the onset of
collapse.  In the analysis, we first give plausible values for $M_{\rm
  out}+M_{\rm eje}$ taking into account the observational results of
electromagnetic counterparts of GW170817.  Then, the energy
conservation law gives a relation between the maximum mass, $M_{\rm
  max}$, and angular momentum of the remnant neutron star at the onset
of collapse. This relation indicates that for smaller values of
$M_{\rm max}$, the collapse occurs at higher angular momentum (i.e.,
at larger values of $f_r$: cf.~Fig.~\ref{fig2}). We also find that the
correlation between $M_{\rm max}$ and $M_f$: the gravitational mass at
the onset of collapse. Thus, for smaller values of $M_{\rm max}$,
$M_f$ has to be larger, i.e., less energy and angular momentum have to
be dissipated prior to the onset of collapse.

We find that the energy conservation laws can be satisfied for a wide
range of equations of state with various values
of $f_r=1.0$--1.2. Also, it is found that the combination of energy and
angular momentum conservation laws gives plausible values of $E_{\rm
  GW,p}$ and $E_\nu$ for the equations of state in which $M_{\rm max}$
is between $\sim 2.1M_\odot$ and $\sim 2.3M_\odot$.  In particular,
the cases of $M_{\rm max} \alt 2.25M_\odot$ result in quite plausible
values of $E_{\rm GW,p}$ and $E_\nu$.  For $M_{\rm
  max}=2.30M_\odot$--$2.35M_\odot$, it is still possible to find
solutions although for such a case, we need to require large energy
dissipation by gravitational radiation, $E_{\rm GW,p} \agt 0.11M_\odot
c^2$.  It is also found that if the value of $M_{\rm max}$ is not very
high, i.e., $M_{\rm max} \leq 2.1M_\odot$, the collapse is likely to
occur before the velocity profile of the remnant neutron star relaxes
to a rigidly rotating state. Thus, we infer that if $M_{\rm max} \alt
2.1M_\odot$, the remnant neutron star collapses to a black hole in the
timescale of angular momentum transport inside it.

The previous analysis often assumes $f_r \approx
1.2$~\cite{MM2017,shibata17,Rezzolla18,Ruiz18}, i.e., the collapse
occurs at a rapidly rotating stage of the remnant neutron star. As
Fig.~\ref{fig2} shows, this assumption together with the energy
conservation law would automatically leads to an inaccurate conclusion
that the value of $M_{\rm max}$ is small, $\alt 2.1M_\odot$ (i.e., the
conclusion is derived from the assumption).  The lesson obtained from
our present analysis is that inappropriate assumptions could lead to
an inaccurate constraint on the maximum mass.

The framework of our analysis for the maximum mass of neutron stars
developed in this paper will be applied for future events, in which
the remnant after the mergers is a massive neutron star that
eventually collapses to a black hole, by simply replacing the values
of $M$ and $M_{\rm out}+M_{\rm eje}$. This analysis will be in
particular interesting if post-merger gravitational waves are observed
in future. For this case, $E_{\rm GW,p}$ and $J_{\rm GW,p}$ will be
constrained, and then, the constraint for the equation of state and
the value of $M_{\rm max}$ will be better imposed. Furthermore, it
will be also feasible to infer how much energy is carried away by
neutrinos.

\acknowledgments

We thank Koutarou Kyutoku for helpful discussions.  This work was in
part supported by Grant-in-Aid for Scientific Research (Grant
Nos.~16H02183 and 18H01213) of Japanese MEXT/JSPS.  We also thank the
participants of a workshop entitled ``Nucleosynthesis and
electromagnetic counterparts of neutron-star mergers'' at Yukawa
Institute for Theoretical Physics, Kyoto University (No. YITP-T-18-06)
for many useful discussions. Numerical computations were performed at
Oakforest-PACS at Information Technology Center of the University of
Tokyo, XC50 at National Astronomical Observatory of Japan, and XC30 at
Yukawa Institute for Theoretical Physics.

\appendix

\section{Model equations of state}

\begin{figure*}[t]
\epsfxsize=3.4in \leavevmode \epsffile{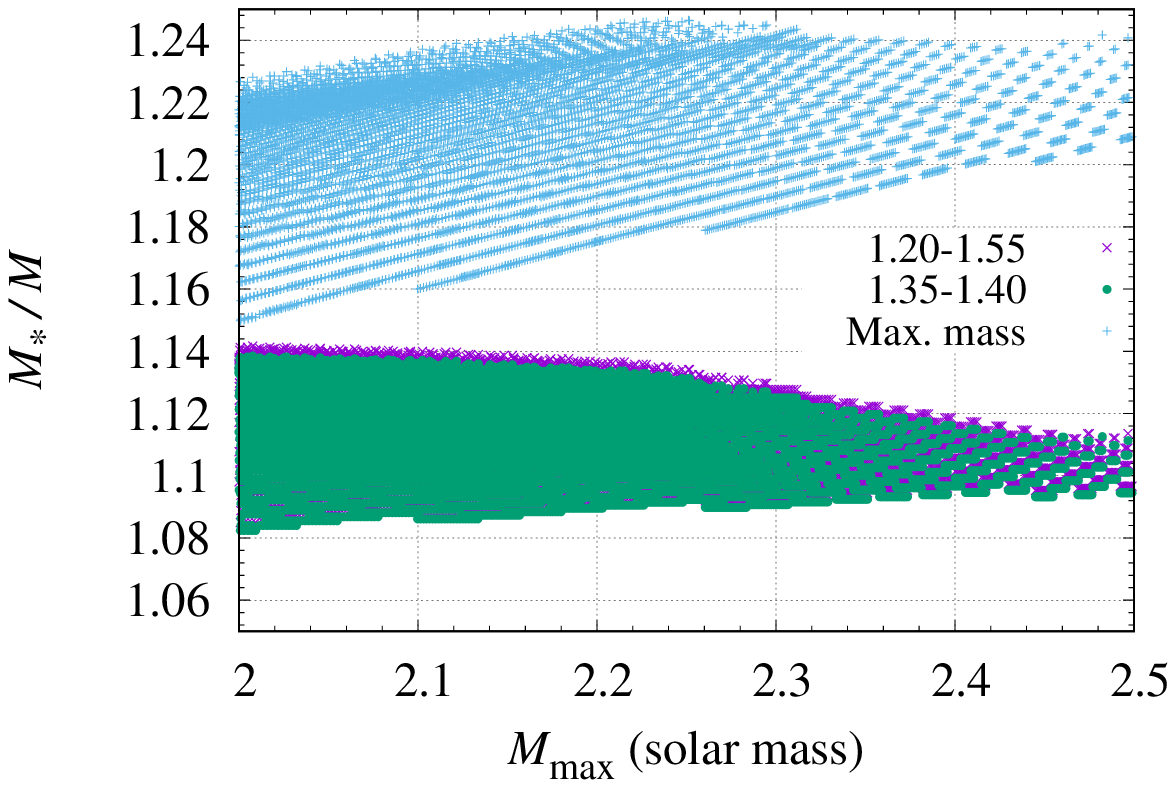}~~
\epsfxsize=3.4in \leavevmode \epsffile{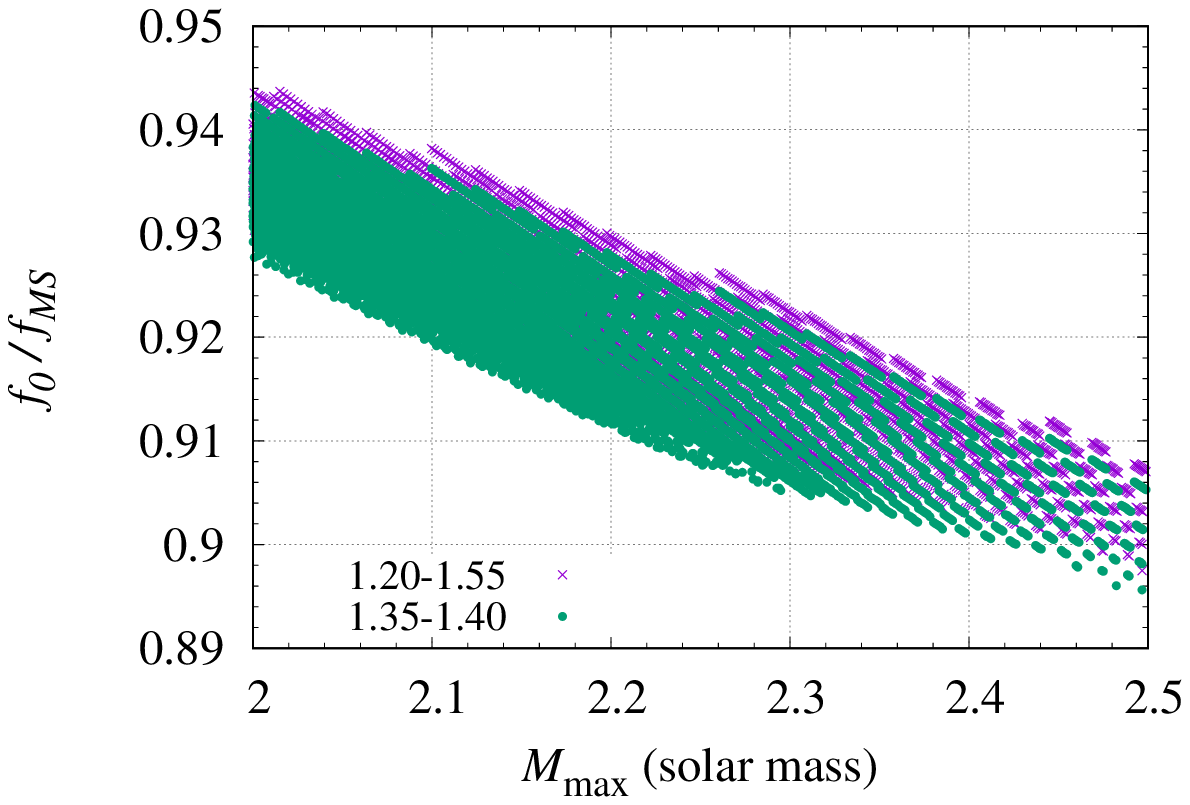}
\caption{Left panel: $f_0$ for the case of $1.20$-$1.55M_\odot$ and
  $1.35$-$1.40M_\odot$ binary neutron stars and $f_{\rm MS}$ for
  spherical maximum-mass stars as a function of the maximum mass for a
  variety of equations of state. Right panel: The same as the left 
  panel but for $f_0/f_{\rm MS}$.  We plot the data with equations of
  state by which $2M_\odot$ neutron stars are reproduced and
  dimensionless tidal deformability of $1.35M_\odot$ star,
  $\Lambda_{1.35}$, is smaller than 1000.
\label{figa1}}
\end{figure*}

To model the neutron-star equation of state with a small number of
parameters, we employ a piecewise polytrope introduced by Read et
al.~\cite{Read09}.  It is a phenomenologically parametrized equation
of state of the form 
\begin{equation}
 P (\rho) = \kappa_i \rho^{\Gamma_i} ~~ {\rm for} ~~ \rho_{i-1} \le
  \rho < \rho_i ~~ (1 \le i \le n) ,
\end{equation}
where $n$ is the number of the pieces used to parametrize an equation
of state, $\rho_i$ is the rest-mass density at the boundary of two
neighboring $i$th and $(i+1)$th pieces, $\kappa_i$ is the polytropic
constant for the $i$th piece, and $\Gamma_i$ is the adiabatic index
for the $i$th piece. Here, $\rho_0 = 0$, and other parameters $(\rho_i
, \kappa_i, \Gamma_i)$ are freely chosen.  Requiring the continuity of
the pressure at each value of $\rho_i$, $2n$ free parameters,
$(\kappa_i,\Gamma_i)$, determine the equation of state completely. The
specific internal energy, $\varepsilon$, is determined by the first
law of thermodynamics and continuity of each variable at $\rho_i$. 

In this paper we employ the case of $n=3$ and fix
$\rho_2=10^{15}\,{\rm g/cm^3}$ (and $\rho_3=\infty$).  The
lowest-density piece models the crust equation of state and the other
two do the core equation of state.  Following Ref.~\cite{Read09}, 
the equation of state for the crust region is fixed by
$\Gamma_1=1.35692395$ and 
$\kappa_1/c^2=3.99873692 \times 10^{-8}\,( {\rm g/cm}^3 )^{1 - \Gamma_1}$.
The equation of state for the core region is determined by two
adiabatic indices, $\Gamma_2$ and $\Gamma_3$, and the pressure, $p$,
at a fiducial density $\rho_f=10^{14.7}\,{\rm g/cm^3}$. Here, $p$ is
closely related to the radius and tidal deformability of neutron
stars~\cite{lattimerprakash2001}.  For $\Gamma_2$, we employ a wide
range of values between 2.10--5.00.  $\Gamma_3$ is chosen to be small
values in a narrow range between 2.00--2.91 because for the large
values, the sound speed exceeds the speed of light even for stable
neutron stars while for the small values, the maximum mass for given
equations of state cannot exceed $2M_\odot$~\cite{two}.  The value of
$\log_{10} p$ is varied between 33.8 and 34.8. Here, for large values
of $\Gamma_2$, only small values of $p$ is accepted (e.g., for
$\Gamma_2=4.0$ and 5.0, $p \alt 10^{34.35}\,{\rm dyn/cm^2}$ and $p
\alt 10^{33.9}\,{\rm dyn/cm^2}$, respectively).  If this condition is
not satisfied, the sound velocity exceeds the speed of light near
$\rho \sim 10^{15}\,{\rm g/cm^3}$. With the given values of
$\Gamma_2$, $\Gamma_3$, and $p$, $\kappa_2$, $\kappa_3$, and $\rho_1$
are determined as
$\kappa_2 = p \rho_{f}^{-\Gamma_2}$, 
$\kappa_3 = \kappa_2 \rho_{2}^{\Gamma_2-\Gamma_3}$, and 
$\rho_1 = ( \kappa_1 / \kappa_2 )^{1 / ( \Gamma_2 - \Gamma_1 )}$. 


We analyzed the values of $f_0$ and $f_{\rm MS}$ for spherical neutron
stars with the piecewise polytropes.  Figure~\ref{figa1} displays a 
result.  In this figure, we plot the data in the piecewise polytropes
with which $2M_\odot$ neutron stars are reproduced and dimensionless
tidal deformability of $1.35M_\odot$ neutron stars satisfies
$\Lambda_{1.35}\leq 1000$ taking into account the results of
GW170817~\cite{GW170817}.  It is found that $f_0$ and $f_{\rm MS}$ are
in a wide range approximately between $1.08$ and $1.14$ and between
1.14 and 1.24, respectively. Because $f_0$ and $f_{\rm MS}$ are weakly
correlated, $f_0/f_{\rm MS}$ is not distributed as widely as $f_{\rm
  MS}$, but it is still in a fairly wide range as
$0.895$--$0.945$. This value depends weakly on each mass of binaries:
For $1.20$--$1.55M_\odot$ binaries, the values of $f_0/f_{\rm MS}$ is
0.2--0.3\% larger than that for $1.35$--$1.40M_\odot$ binaries for a
given equation of state. 

One interesting and key fact for the context of the present paper is
that $f_0/f_{\rm MS}$ is correlated with $M_{\rm max}$; for larger
value of $f_0/f_{\rm MS}$, the maximum mass of neutron stars is
smaller.  The reason for this is that for the equations of state with
smaller maximum mass, the neutron star at the maximum mass is less
compact and thus has smaller value of $f_{\rm MS}$.  This fact implies
that for smaller maximum mass, the mass of the remnant neutron-star
mass at the onset of collapse, $M_f=M f_0/f_{\rm MS}-(M_{\rm
  out}+M_{\rm eje})/f_{\rm MS}$, is larger (see Eq.~(\ref{eq1.3})),
and hence, the merger remnant is more subject to gravitational
collapse.  We note that the upper boundary of $f_0/f_{\rm MS}$ for a
give value of $M_{\rm max}$ is determined by the condition of
$\Lambda_{1.35} \leq 1000$.


\end{document}